\documentclass[twocolumn]{aastex7}

\usepackage{amsmath}
\usepackage{wasysym}
\usepackage{savesym}
\savesymbol{tablenum}
\usepackage{siunitx}
\restoresymbol{SIX}{tablenum}
\usepackage{booktabs}
\usepackage{rotating}
\usepackage[T1]{fontenc}
\usepackage{blindtext}

\newcommand{\upd}{\mathrm{d}}

\DeclareSIUnit\Mearth{M_\oplus}
\DeclareSIUnit\Mjupiter{M_J}
\DeclareSIUnit\Rearth{R_\oplus}
\DeclareSIUnit\Rjupiter{R_J}
\DeclareSIUnit\Mmoon{M_\textrm{\leftmoon}}
\DeclareSIUnit\JEM{J_{EM}}

\graphicspath{{./}{figures/}}

\received{14.11.2025}
\revised{03.03.2026}
\accepted{04.03.2026}
\submitjournal{APJ}

\begin{document}

\title{Smoothed Particle Hydrodynamics in \texttt{pkdgrav3} for Shock Physics Simulations. I. Hydrodynamics}

\author[orcid=0000-0001-9682-8563,gname='Thomas', sname='Meier']{Thomas Meier} 
\affiliation{Department of Astrophysics, University of Zurich, Winterthurerstrasse 190, CH-8057 Zurich, Switzerland}
\email[show]{thomas.meier5@uzh.ch}

\author[orcid=0000-0002-0757-5195,gname='Douglas', sname='Potter']{Douglas Potter} 
\affiliation{Department of Astrophysics, University of Zurich, Winterthurerstrasse 190, CH-8057 Zurich, Switzerland}
\email{douglas.potter@uzh.ch}

\author[orcid=0000-0002-4535-3956,gname='Christian' ,sname='Reinhardt']{Christian Reinhardt}
\affiliation{Department of Astrophysics, University of Zurich, Winterthurerstrasse 190, CH-8057 Zurich, Switzerland}
\affiliation{Physics Institute, Space Research and Planetary Sciences, Center for Space and Habitability, University of Bern, Sidlerstrasse 5, CH-3012 Bern, Switzerland}
\email{christian.reinhardt@uzh.ch}

\author[orcid=0000-0001-7565-8622,gname='Joachim', sname='Stadel']{Joachim Stadel} 
\affiliation{Department of Astrophysics, University of Zurich, Winterthurerstrasse 190, CH-8057 Zurich, Switzerland}
\email{joachimgerhard.stadel@uzh.ch}

\begin{abstract}
We present \texttt{pkdgrav3}, a high-performance, fully parallel tree-SPH code designed for large-scale hydrodynamic simulations including self-gravity. Building upon the long development history of \texttt{pkdgrav}, the code combines an efficient hierarchical tree algorithm for gravity and neighbor finding with a modern implementation of Smoothed Particle Hydrodynamics (SPH) optimized for massively parallel hybrid CPU/GPU architectures. Its hybrid shared/distributed memory model, combined with an asynchronous communication scheme, allows \texttt{pkdgrav3} to scale efficiently to thousands of CPU cores and GPUs. We validate the numerical accuracy of \texttt{pkdgrav3} using a suite of standard tests, demonstrating excellent agreement with analytic or reference solutions. The code was already used in several peer-reviewed publications to model planetary-scale impacts, where SPH's Lagrangian nature allows accurate tracking of material origin and thermodynamic evolution. These examples highlight \texttt{pkdgrav3}'s robustness and efficiency in simulating highly dynamical, self-gravitating systems. \texttt{pkdgrav3} thus provides a powerful, flexible, and scalable platform for astrophysical and planetary applications, capable of exploiting the full potential of modern heterogeneous high-performance computing systems.
\end{abstract}



\section{Introduction}
Hydrodynamical simulations play a key role across many branches of astrophysics and cosmology, including large-scale structure and galaxy formation \citep{vogelsbergerIntroducingIllustrisProject2014,vogelsbergerCosmologicalSimulationsGalaxy2020}, star formation \citep{teyssierNumericalMethodsSimulating2019}, accretion disks \citep{davisMagnetohydrodynamicsSimulationsActive2020}, and planetary impacts \citep{asphaugGlobalScaleImpacts2015,gabrielRoleGiantImpacts2023}. To accurately model these diverse phenomena, a variety of numerical techniques have been developed to solve the equations of fluid dynamics. They can broadly be categorized into grid-based and meshless approaches, the latter being primarily represented by particle-based methods. 

Mesh-based schemes partition the computational domain into discrete volume elements (cells), whose hydrodynamic state is evolved in time. These cells are typically arranged in a structured mesh that can be adaptively refined in regions requiring higher resolution. This structured layout enables efficient memory access and computational performance \citep{teyssierCosmologicalHydrodynamicsAdaptive2002,bryanENZOADAPTIVEMESH2014}. However, determining where and how to refine the mesh is a non-trivial problem that often requires case-specific fine-tuning. Traditional mesh-based methods are also not Galilean invariant, which can introduce advection errors when simulated structures move relative to the static grid \citep{agertzFundamentalDifferencesSPH2007,wadsleyGasoline2ModernSmoothed2017}. These issues can be mitigated, though not entirely eliminated, by adopting a coordinate system tailored to the problem under investigation or by employing moving-mesh formulations \citep{springelPurSiMuove2010}.

Particle-based methods represent fluids as a set of discrete particles that move with the local flow. Each particle carries the hydrodynamic quantities of a fluid element, which are evolved according to the equations of motion. The most widely used particle-based approach is smoothed particle hydrodynamics (SPH) \citep{gingoldSmoothedParticleHydrodynamics1977,lucyNumericalApproachTesting1977}, in which continuous fields are reconstructed by kernel interpolation over neighboring particles. This fully Lagrangian formulation provides several advantages over mesh-based schemes. Because particles follow the motion of the fluid, SPH is naturally adaptive: resolution automatically increases in regions of high density or strong compression without the need for an explicit refinement criterion \citep[see reviews in][]{rosswogAstrophysicalSmoothParticle2009,priceSmoothedParticleHydrodynamics2012}. The method also conserves mass and momentum by construction and is free of advection errors, since the computational elements move with the flow. However, standard SPH formulations have well-known shortcomings. Artificial viscosity is required to capture shocks, which can introduce excessive dissipation in regions of smooth flow \citep{monaghanSmoothedParticleHydrodynamics1992,monaghanSmoothedParticleHydrodynamics2005}. Furthermore, traditional density-entropy formulations suppress fluid mixing and hydrodynamic instabilities at contact discontinuities due to spurious surface tension effects \citep{agertzFundamentalDifferencesSPH2007, reinhardtBifurcationHistoryUranus2020,ruiz-bonillaDealingDensityDiscontinuities2022}. Several improved variants have been developed to address these issues, including pressure-based and geometric-density SPH formulations \citep{readResolvingMixingSmoothed2010,hopkinsNewClassAccurate2015,borrowSphenixSmoothedParticle2021}, as well as schemes that employ higher-order kernels or adaptive viscosity switches \citep{cullenInviscidSmoothedParticle2010,priceSmoothedParticleHydrodynamics2012}.

Thanks to its fully Lagrangian, adaptive nature, SPH is particularly well-suited for simulating highly dynamic, strongly deforming flows. Its ability to naturally follow fluid motion and resolve fine structures without an explicit mesh refinement criterion makes it ideal for astrophysical events involving large density contrasts, free surfaces, and complex geometries. These properties have made SPH a method of choice for modeling planetary-scale collisions, where tracking the motion and mixing of individual material parcels is essential \citep[e.g.,][]{canupSimulationsLateLunarforming2004,asphaugMercuryOtherIronrich2014,emsenhuberSPHCalculationsMarsscale2018, chauFormingMercuryGiant2018, reinhardtBifurcationHistoryUranus2020,reinhardtFormingIronrichPlanets2022,timpeSystematicSurveyMoonforming2023,ballantyneInvestigatingFeasibilityImpactinduced2023,meierSystematicSurveyMoonforming2024,ballantyneSputnikPlanitiaImpactor2024}. This capability is crucial for addressing questions such as tracing the origin of material, following the fate of core material, and understanding the formation of moons or debris disks resulting from giant impacts.

For decades, leading simulations in these fields have relied on high-performance computing (HPC) facilities equipped with powerful supercomputers. Efficient use of such systems requires codes to scale across multiple nodes and to exploit many CPU cores per node. More recently, the increasing prevalence of GPU-accelerated systems has added another layer of complexity: simulation codes must leverage GPUs to fully utilize modern HPC resources and remain competitive for computational grants. A prominent example is the $N$-body code \texttt{pkdgrav3}, the successor to the widely used \texttt{pkdgrav} code. Building on the excellent scaling and parallel efficiency of its predecessor, \texttt{pkdgrav3} is capable of simulating trillions of particles across thousands of CPU and GPU nodes \citep{potterPKDGRAV3TrillionParticle2017}, making it a leading platform for large-scale gravitational simulations. Recently, a hydrodynamics module to simulate galaxy formation based on the Meshless Finite Mass method (MFM, \citealt{hopkinsNewClassAccurate2015}) was implemented in the code by \citet{alonsoasensioMeshfreeHydrodynamicsPkdgrav32023}.

In this work, we implement an independent SPH module in \texttt{pkdgrav3}, emphasizing high-performance algorithms and GPU acceleration to efficiently exploit modern heterogeneous supercomputing architectures. This addition enables fully coupled simulations of self-gravity and hydrodynamics within a single, scalable framework to achieve exceptional performance and unprecedented resolution.

We use a standard SPH formulation where the equations of motion are derived self-consistently from the Lagrangian following \citet{springelCosmologicalSmoothedParticle2002}. We chose the usual (standard) density-energy formulation \citep{monaghanSmoothedParticleHydrodynamics1992} because these quantities are readily available for common EOS used for modeling planetary collisions. In the literature other flavors of SPH, such as DISPH \citep{saitohDENSITYINDEPENDENTFORMULATIONSMOOTHED2013,hopkinsGeneralClassLagrangian2013} or GDF \citep{wadsleyGasoline2ModernSmoothed2017} have been proposed to improve the method at modeling contact discontinuities and improve fluid mixing. However, these either struggle with capturing shocks, modeling free surfaces, require thermodynamic quantities not provided by all EOS, or cannot be derived from a Lagrangian requiring ad-hoc choices when disretizing the fluid equations of motion. In our implementation we use the interface/free-surface correction proposed in \citet{ruiz-bonillaDealingDensityDiscontinuities2022} and enforce entropy conservation in adiabatic flows without explicitly requiring entropy information \citep{reinhardtNumericalAspectsGiant2017} making it a robust and well-suited method for shock-physics modeling.

This paper is structured as follows. In Section~\ref{sec:Code_Description}, we describe the \texttt{pkdgrav3} $N$-body code and the details of our SPH implementation based on it. In Section~\ref{sec:Hydrodynamic_tests}, we present a large suite of hydrodynamic tests. In Section~\ref{sec:Scaling_tests} we show the performance and scaling of the code. We give a summary and final remarks in Section~\ref{sec:Conclusions}.

\section{Code Description}\label{sec:Code_Description}
The $N$-body code \texttt{pkdgrav} was first introduced by \citet{stadelCosmologicalNbodySimulations2001}. Since then, it has been used in a variety of numerical studies: it has been directly compared with other $N$-body simulation codes \citep{powerInnerStructureLCDM2003,diemandConvergenceScatterCluster2004}; adapted for simulations of planetesimal dynamics using the hard- and soft-sphere discrete element method (DEM) \citep{richardsonDirectLargeScaleNBody2000,schwartzImplementationSoftsphereDiscrete2012,marohnicEfficientNumericalApproach2023}; and served as the foundation for the development of the SPH codes \texttt{GASOLINE} \citep{wadsleyGasolineFlexibleParallel2004} and its successor \texttt{GASOLINE2} \citep{wadsleyGasoline2ModernSmoothed2017}.

Since its initial development, the pure $N$-body version of \texttt{pkdgrav} was completely rewritten into \texttt{pkdgrav2} which implemented a Fast Multipole Method (FMM) algorithm, bringing the gravity calculation scaling from $\mathcal{O}(N\log N)$ to $\mathcal{O}(N)$, as well as more sophisticated time-stepping criteria based on a better estimate of dynamical time in $N$-body simulations (see \citet{zempOptimumTimesteppingScheme2007}, \citet{diemandClumpsStreamsLocal2008} and \citet{stadelQuantifyingHeartDarkness2009}). The code was further developed into \texttt{pkdgrav3} to take full advantage of multinode and multicore architectures using hybrid MPI/pthreads support as well as GPU acceleration using NVidia CUDA programming \citep{potterPKDGRAV3TrillionParticle2017}. Furthermore, large parts of the tree walk and the interactions were vectorized on the CPU, using SIMD, further increasing performance of \texttt{pkdgrav3}. This version of the code has also been compared to other $N$-body codes \citep{schneiderMatterPowerSpectrum2016,garrisonHighfidelityRealizationEuclid2019} and one of the largest simulations of the large-scale structure of the Universe to date has been performed with \texttt{pkdgrav3} for the Euclid Collaboration \citep{potterPKDGRAV3TrillionParticle2017,knabenhansEuclidPreparationII2019}.

We implement SPH on this foundation, reusing as much of the improvements as possible to get a high-performance SPH code for the application in general astrophysical simulations and specifically to simulate impact processes on planetary scales. The source code of \texttt{pkdgrav3}, including the SPH implementation, is released under the GNU General Public License (version 3, GPLv3)\footnote{\url{https://www.gnu.org/licenses/gpl-3.0.en.html}} on Bitbucket\footnote{\url{https://bitbucket.org/dpotter/pkdgrav3}}, with the current version deposited to Zenodo: \href{https://zenodo.org/records/18754678}{10.5281/zenodo.18754678}. Documentation is available\footnote{\url{https://pkdgrav3.readthedocs.io/}}.

In this section, we first describe the general architecture of the \texttt{pkdgrav3} code, then we outline the FMM gravity solver, with emphasis on the parts that will be reused by the SPH solver, and finally describe the SPH solver itself.

\subsection{Code architecture}
\texttt{pkdgrav3} inherits its parallelization strategy from its predecessor \texttt{pkdgrav} \citep{stadelCosmologicalNbodySimulations2001} and it is described in more detail in \citep{potterPKDGRAV3TrillionParticle2017}. The code is structured into four abstraction layers. This design ensures a clean separation between communication and computation, allowing developers to focus on algorithmic implementation without having to deal with communication details or patterns. The layered structure also provides high flexibility and portability across system architectures, cluster topologies, and scientific applications. The four abstraction layers are described below:

\begin{enumerate}
    \item \textbf{Master Layer (MSR)}: Executed serially on a single process, this layer defines the program's overall workflow - from reading input parameters to distributing computational tasks to other processes through the next layer.
    \item \textbf{Processor Set Tree (PST)}: This layer organizes processes into a binary tree, with the master process as the root. When a computation is dispatched to the PST, each node propagates the relevant information to its child nodes recursively, then begins its own portion of the computation, which is carried out by the PKD layer.
    \item \textbf{PKD Layer}: The shallowest layer, and the only one permitted to modify particle and tree data. It runs on all cores and contains all the physics modules. The PKD layer consists of quasi-serial code that operates on data local to each core. Since most algorithms (e.g., force calculations) require access to ``remote'' data from other cores, this layer uses the Machine-Dependent Layer (MDL) to fetch such data on demand. This mechanism closely resembles a Partitioned Global Address Space (PGAS) model, where data is exposed as a virtual vector and remote elements can be accessed by index.
    \item \textbf{Machine-Dependent Layer (MDL)}: The lowest layer, responsible for communication, task management, and scheduling. It maintains caches of frequently accessed data from other processes to minimize communication. MDL supports both shared-memory (via boost threads) and distributed-memory (via MPI) parallelism, or a hybrid of the two. It also handles GPU offloading when available.
\end{enumerate}

The simulation domain, populated with particles, is decomposed into a binary tree of rectangular cuboid subdomains using Orthogonal Recursive Bisection (ORB). This decomposition aligns with the Processor Set Tree (PST) structure and is performed such that all leaf domains contain an equal number of particles, ensuring perfect memory balance. The particles within each leaf domain are then moved to the processor corresponding to that domain's position in the tree. Each processor subsequently constructs a binary tree of the particles within its assigned subdomain. The result is a global binary tree encompassing all particles: the upper portion of the tree is represented by the PST, while the lower portions correspond to the per-CPU particle trees.

Each processor stores its particles in a contiguous memory block organized as an Array of Structures (AoS). The particle structure consists of a header followed by all the required fields, which are determined at runtime based on the user-specified options. During tree construction, the particle storage is reordered so that, at each level of the tree, particles belonging to the same cell are contiguous in memory. Each tree cell holds pointers into this particle array that define the range of particles belonging to it, specifically, \texttt{pLower} and \texttt{pUpper} indicate the start and end of the corresponding memory region. This layout enables efficient particle loops with minimal cache misses.

\texttt{pkdgrav3} employs three nested levels of parallelism. Distributed-memory parallelism is achieved via MPI, and shared-memory parallelism is handled through boost threads, both managed by the Machine-Dependent Layer (MDL, see above). On each physical compute node, one or more MPI ranks are launched, each utilizing multiple boost threads. The third level of parallelism is data-level parallelism, where particle interactions (and opening criteria) are computed using SIMD (Single Instruction Multiple Data) vectorization on CPUs, using Streaming SIMD Extensions (SSE) and Advanced Vector Extensions (AVX) variants on x86\_64 or NEON on ARM. The build system automatically detects and enables the highest SIMD instruction set supported by the target architecture. On GPUs, data-level parallelism is achieved through SIMT (Single Instruction Multiple Threads) execution. To enable this efficiently, data are converted from the AoS layout to an Array of Structures of Arrays (AoSoA) format during the assembly of the interaction lists (see Sections~\ref{sec:Gravity} and~\ref{sec:Neighbor_search}). Each inner SoA block, referred to as a tile, contains arrays of 32 floating-point values, matching the CUDA warp size.

\subsection{Gravity}\label{sec:Gravity}
\texttt{pkdgrav3} computes the accelerations arising from self-gravity using the Fast Multipole Method (FMM) \citep{greengardFastAlgorithmParticle1997}. The FMM extends the classic Barnes-Hut tree code, with both methods employing direct particle-particle (PP) interactions for nearby particles and a multipole expansion of the mass distribution within distant cells to approximate their gravitational influence (particle-cell, or PC, interactions). The key enhancement introduced by FMM is the inclusion of cell-cell (CC) and cell-particle (CP) interactions, which approximate the gravitational potential field within a ``sink'' cell as induced by the multipole moments of a sufficiently distant ``source'' cell or a single source particle in the case of CP interactions.

The efficiency of a tree-based code, especially when GPU acceleration is involved, depends critically on how interaction lists (PP, PC, CP and CC) are constructed. These lists define which interactions must be evaluated to compute the forces on particles. To build them, the tree is traversed in a node-left-right recursive order over the sink cells, while a checklist of source cells is maintained. For each sink cell, source cells from the checklist are evaluated against an opening criterion, determining whether a source cell should be:

\begin{enumerate}
    \item opened (removed from the checklist and replaced by its children),
    \item added to one of the four interaction lists (depending on its distance from the sink cell),
    \item retained on the checklist for further consideration by the sink cell's descendants during deeper levels of the tree traversal.
\end{enumerate}

Evaluating the opening criterion is a purely arithmetic operation, implemented using SIMD vectorization, and incurs only a negligible computational cost (about $\sim\SI{2}{\percent}$ of the total runtime). The tree traversal begins with the sink cell set to the root of the local processor's tree, while the checklist initially contains the global root cell of the entire simulation domain along with its surrounding periodic replicas. Long-range periodic forces are evaluated using Ewald summation \citep{hernquistApplicationEwaldMethod1991,klessenGRAPESPHFullyPeriodic1997,stadelCosmologicalNbodySimulations2001}

The opening criterion plays a decisive role in controlling both the magnitude and the spatial correlation of the resulting force errors. A detailed discussion of the criterion and its implementation in \texttt{pkdgrav3} can be found in \citet{potterPKDGRAV3TrillionParticle2017}. During tree traversal, the source tree is walked down to the level of individual buckets, which by default contain 16 particles each. On the sink side, the tree is traversed only down to the level of groups, which by default contain 64 particles on CPUs and 256 particles when GPU acceleration is used. When a bucket is encountered during the traversal for a given group, all particles within that bucket are added to the particle-particle (PP) interaction list being assembled for that group.

The resulting interaction lists are then processed in tiles, which are evaluated either on a GPU, using SIMT parallelism, or on the CPU using SIMD vectorization. The contributions from all tiles are accumulated incrementally for the entire group. Once the final tile has been processed, the total accelerations for all particles in the group are obtained. At that point, a leapfrog kick is applied to update the particle velocities (see Section~\ref{sec:Time_Integration}).

\subsection{Smoothed Particle Hydrodynamics}
The equations governing the evolution of an inviscid fluid are called the Euler equations, which express mass, momentum and energy conservation. In Lagrangian form, they are given as

\begin{align}
\frac{\upd\rho}{\upd t}&=-\rho\nabla\cdot \vec{v}\,,\label{eq:Mass_conservation}\\
\rho\frac{\upd \vec{v}}{\upd t}&=-\nabla P+\rho \vec{g}\,,\label{eq:Momentum_conservation}\\
\rho\frac{\upd u}{\upd t}&=-P\nabla\cdot \vec{v}\,,\label{eq:Energy_conservation}
\end{align}

\noindent where $\rho$, $P$, $\vec{v}$ and $u$ are the fluid density, pressure, velocity and internal energy respectively, while $\vec{g}$ is the force due to external and self-gravity. To close the system, an equation of state (EOS) is needed. We discuss the currently available EOSs in Section~\ref{sec:EOS}.

SPH is based on local summations over neighbors. For a fluid quantity $f_j$ that is known at particle positions $x_j$, a smoothed value at any position $\vec{x}_i$ can be calculated as

\begin{equation}
f(\vec{x}_i)=\sum\limits_{j}f_jW(\vec{x}_i-\vec{x}_j,h_j)\,,
\end{equation}

\noindent where $W$ is the kernel function and $h_j$ is the smoothing length. \texttt{pkdgrav3} provides the cubic spline kernel function \citep{monaghanSmoothedParticleHydrodynamics1992} and the Wendland C2, C4 and C6 kernel function variants \citep{dehnenImprovingConvergenceSmoothed2012}. There are several ways to discretize the Euler equations using kernel summations. We adopt the formulations derived from the Lagrangian, as they naturally incorporate corrections arising from a variable smoothing length \citep{springelCosmologicalSmoothedParticle2002,priceSmoothedParticleHydrodynamics2012}. To evaluate these local summations, we first have to find the neighbors to sum over. We describe how we do this and how we then calculate density estimates and time derivatives and integrate in time in the next sections.

\subsubsection{Equation of state}\label{sec:EOS}
An equation of state (EOS) is a relation between thermodynamic quantities, describing a specific material. The relation between the pressure and the internal energy and the density $P(\rho,u)$ is used to close the Euler equations, while other values provided by the EOS (e.g., the sound speed, converting internal energy in temperature and vice-versa) are used in other places of the SPH formalism. \texttt{pkdgrav3} provides a built-in ideal gas EOS implementation. If available at compile-time, the generalized EOS interface \texttt{EOSlib} \citep{meierEOSResolutionConspiracy2021,meierEOSlib2021} can be included to make use of all the EOS supported by it. The currently available modules for \texttt{EOSlib} are the Tillotson EOS \citep{tillotsonMetallicEquationsState1962}, tabulated EOS created from ANEOS and M-ANEOS input decks \citep{thompsonImprovementsCHARTRadiationhydrodynamic1974,meloshHydrocodeEquationState2007,thompsonMANEOS2019,meierANEOSmaterial2021} and tabulated EOS for hydrogen, helium and H-He mixtures based on REOS.3 \citep{beckerINITIOEQUATIONSSTATE2014,wooDidUranusRegular2022} and the SCvH EOS \citep{saumonEquationStateLowMass1995,vazanEffectCompositionEvolution2013,matzkevichOutcomeCollisionsGaseous2024}, but due to the modular nature of \texttt{EOSlib}, including new EOSs is relatively simple.

\subsubsection{Neighbor search}\label{sec:Neighbor_search}
\begin{figure*}[ht!]
\centering
\includegraphics[width=0.3\linewidth]{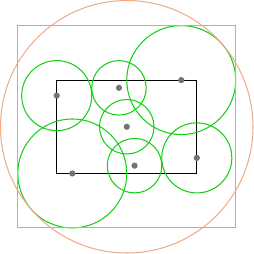}\hfill\includegraphics[width=0.3\linewidth]{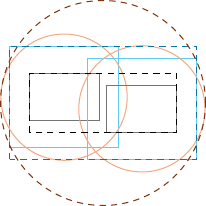}\hfill\includegraphics[width=0.36\linewidth]{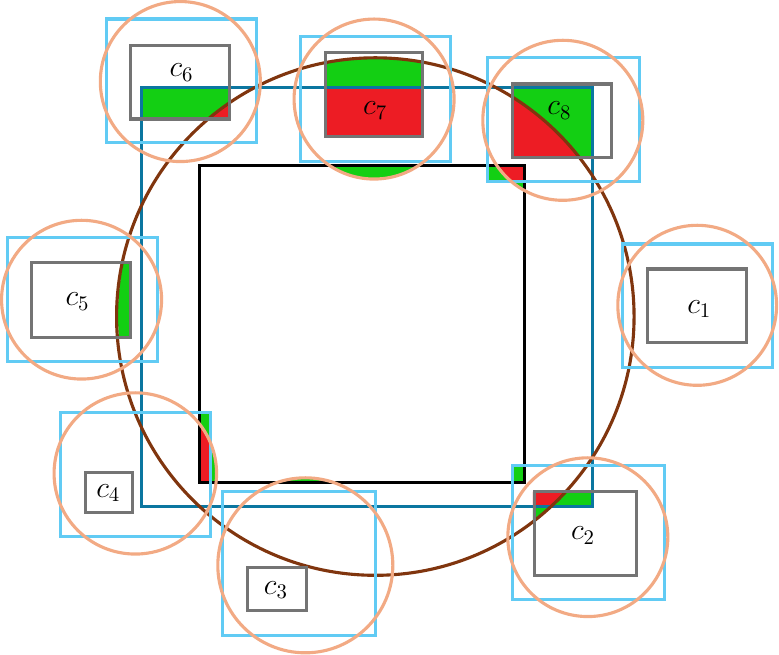}
\caption{Left panel: 2D illustration of the particles (gray dots) in the cell bound (black) with their kernel balls (green) and the resulting ball bounds, the box-of-balls in blue and the ball-of-balls in orange. Center panel: combining two cells in the tree build results in combined bounds. The original cell bounds are gray solid lines, while the combined cell bound is shown by a black dashed line. The combination of the boxes of balls (solid light blue lines) is shown as a dashed dark blue line, while the combination of the balls of balls (solid orange lines) is shown as a dashed brown line. Right panel: during the tree walk, the bounds are compared with each other. The bounds of the group for which the tree is currently walked (sink), are shown in black, dark blue and brown, while the gray, light blue and orange colored bounds represent different cells for which their fate is calculated (source). Depending on the overlap, cells are either opened or rejected. An overlap is only counted if both the box-of-balls and ball-of-balls overlap. Regions with only one overlap are colored green, and those with two overlaps are colored red. Because the cells $c_1, c_3$ and $c_5$ have no red regions, they are rejected in all cases. If only gather neighbors are needed, only the cells $c_2, c_6, c_7$ and $c_8$ have red regions in their cell bound and are accepted while the cell $c_4$ is rejected. If both gather and scatter neighbors are needed, the cell $c_4$ is also accepted because it has a red overlap with the cell bound of the sink.}
\label{fig:Ball_Bounds_Opening}
\end{figure*}

Finding the neighbors of a particle has always been one of the conceptually most complex parts of the SPH method. Early implementations of SPH employed the Gaussian as a kernel function, making every particle a neighbor of every other particle \citep{gingoldSmoothedParticleHydrodynamics1977,lucyNumericalApproachTesting1977}. This approach is inherently of order $\mathcal{O}\bigl(N^2\bigr)$ and therefore becomes computationally infeasible for larger particle numbers. To address this, kernel functions with compact support were introduced, which limit the number of interacting neighbors \citep{monaghanParticleMethodsHydrodynamics1985} and naturally enable an adaptive spatial resolution by maintaining a constant neighbor count. However, even with compact kernels, the naive search for neighbors remains an $\mathcal{O}\bigl(N^2\bigr)$ operation. This problem can be solved by constructing an auxiliary data structure, for example a tree \citep[e.g.,][]{hernquistTREESPHUnificationSPH1989,wadsleyGasolineFlexibleParallel2004,wadsleyGasoline2ModernSmoothed2017} or a grid \citep[e.g.,][]{schallerSwiftModernHighly2024}. If one constructs a tree, each tree node can contain a precomputed bound of the particles in the node. By walking the tree for each particle, assembling the list of neighbors in a priority queue, large parts of the domain can be rejected by checking the bounds of the tree nodes. By cleverly choosing the way the tree is walked, and by reusing the priority queue from one particle to the next, only recomputing the distances, even larger parts of the tree can be ignored.

We improve on this idea by using the binary tree that underlies the FMM gravity solver in \texttt{pkdgrav3} for the neighbor search. Each tree cell contains a coordinate-aligned bounding box enclosing the particle coordinates (the \emph{cell bound}) and a \emph{ball bound} object. The size of the latter is determined from the kernel ball sizes of the particles (see also \citet{gaftonFastRecursiveCoordinate2011} for a similar approach). The ball bound object stores two types of geometric enclosures (see the left panel of Figure~\ref{fig:Ball_Bounds_Opening}): (1) a coordinate-aligned bounding box, and (2) an enclosing sphere. While the tree itself is constructed in a top-down manner, the bounds of each cell are computed bottom-up. Starting from the leaf cells (also called buckets), the bounds of parent cells are obtained by recursively combining those of their children (see center panel of Figure~\ref{fig:Ball_Bounds_Opening}). At each combination step, both the box and the sphere are updated to enclose those of the child cells. The resulting box-of-balls is guaranteed to be the smallest coordinate-aligned box that encloses all contained balls. For the ball-of-balls, however, we only ensure that it fully encloses all constituent balls, without requiring it to be the smallest possible. Computing the exact smallest enclosing ball of a set of balls is computationally expensive \citep{cavaleiroDualSimplextypeAlgorithm2021,fischerSmallestEnclosingBall2003} and offers little advantage for our purposes.

To now find the neighbors using this tree, we employ a modified version of \texttt{pkdgrav3}'s FMM opening criterion, treating SPH neighbors as particle-particle interactions (see Section~\ref{sec:Gravity}). For each group of particles (called sink, which is the smallest cell containing at least 256 particles by default), an interaction list for particle-particle interactions is assembled by walking the tree, calling upon the opening criterion to decide for each cell (source) if it should be opened or rejected (or treated as particle-cell or cell-cell interactions for gravity). The evaluation of the opening criterion is done vectorized for multiple cells at a time. Both the enclosing box-of-balls and ball-of-balls of the sink cell are compared to the cell bound of a potential source cell to check if it can possibly have any particles inside the kernel of any particle in the sink cell. If this is true for \emph{both enclosing shapes}, the source cell is opened or, if it is a bucket, the particles are potential gather neighbors and are thus added to the interaction list. If scatter neighbors are also needed (in the case of the pass that calculates the time derivatives) the same procedure is employed by comparing the enclosing shapes of the source cell to the cell bound of the sink cell, again adding identified potential neighbors to the interaction list. The four possible results of this intersection check are: no intersection, potential gather neighbor, potential scatter neighbor and potential gather/scatter neighbor. Some possible arrangements of this are shown in the right panel of Figure~\ref{fig:Ball_Bounds_Opening}. Performing these overlap tests for a full group (256 particles by default) means that we need only one tree walk per group, instead of one per particle, which significantly reduces the amount of work and communication, contributing to the good performance of the \texttt{pkdgrav3} code.

The above procedure ensures that any particle that could conceivably be a neighbor of any particle in the group ends up on the PP-interaction list. But it also contains many particles that are not in fact neighbors of any sink particle, with typical lists containing around $10-100\, N_{neigh}$ particles. However, trying to reduce the interaction list at this point, by checking particle pairs between the sink group and the interaction list, is inefficient. Instead the information of the particles in the sink group and the interaction list are packaged together and sent to have their interactions evaluated. This is done vectorized, either on the CPU or the GPU, including the selection of the actual neighbors for each particle from the interaction list.

Typically, one expects runtime to increase with the number of neighbors, since the computational cost scales as $\mathcal{O}\bigl(N_{neigh}\bigr)$, while the number of substeps decreases only as $\mathcal{O}\bigl(N_{neigh}^{1/3}\bigr)$ due to the larger smoothing length $h$. However, all tests performed with varying $N_{neigh}$ in Section~\ref{sec:Hydrodynamic_tests} show a speedup of about \SI{35}{\percent} when increasing $N_{neigh}$ from 32 to 400. The time per substep increases by roughly \SI{40}{\percent}, consistent with interaction lists that contain 2.3 times as many particles. Thus, the size of the interaction lists does not scale linearly with $N_{neigh}$.

At extremely steep density gradients, such as at the interface between a rocky planet's surface and a surrounding vapor cloud generated by an impact, the neighbor selection procedure can produce very large interaction lists. Although this affects only a small number of groups, in multi-node simulations the resulting communication overhead can slow down the entire computation. To mitigate this effect, the group and bucket sizes are adjusted dynamically based on the interaction list size from the previous step, in the most extreme cases reducing the bucket size to a single particle. This approach substantially decreases interaction list sizes in regions with sharp density contrasts, resulting in markedly improved performance.

\subsubsection{Density and smoothing length calculation}\label{Density_and_smoothing_length_calculation}
As the density is needed to calculate the time derivatives (see Section~\ref{sec:Time_derivatives}), it has to be updated first. Thus, the first pass (loop over the particles) of each time step is to calculate the density for each particle $i$ as the sum of the masses of all neighboring particles $j$ weighted with the kernel function

\begin{equation}\label{eq:Density_estimate}
\rho_i = \sum_jm_jW_{ij}(h_i)\,,
\end{equation}

\noindent where we define $W_{ij}(h_x) = W(\left\vert\vec{x}_i-\vec{x}_j\right\vert,h_x)$. 

The summation is computed using the interaction list assembled during the neighbor search (see Section~\ref{sec:Neighbor_search}) and is performed vectorized either on the CPU or the GPU. For each particle in the group and each potential neighbor, the distance is calculated and depending if this is smaller than the kernel size, the particle is treated as a neighbor, otherwise it is skipped.

The smoothing length $h_i$ is apriori unknown, but the derivation from the Lagrangian provides a natural criterion for automatically adapting it to the local particle distribution \citep[see][]{springelCosmologicalSmoothedParticle2002,priceSmoothedParticleHydrodynamics2012,hopkinsGeneralClassLagrangian2013}, leading to a smooth spatial variation of $h$. To achieve this, we compute the density according to Equation~\eqref{eq:Density_estimate} and its derivative with respect to $h_i$, which are then used in Newton-Raphson iterations to maintain a constant effective mass in the kernel:

\begin{equation}
M_{tot}^i=\int_{V_i}\rho_i \upd V \approx \frac{4}{3}\pi R_{kernel}^3(h_i)\rho_i\,.
\end{equation}

We do not have to walk the tree in each iteration to find the neighbors, because we keep the precomputed interaction list. To accommodate expanding kernels we slightly over-select potential neighbors during the neighbor search for the density pass by assuming a slightly larger kernel. After the density pass, the ball bounds objects (see Section~\ref{sec:Neighbor_search}) in the tree are updated with the new value to give subsequent passes accurate bounds. In addition to the density, this pass also calculates the correction term (also called $\nabla h$ term) needed to account for the gradient of the smoothing length given by

\begin{equation}
\Omega_i = 1 + \frac{h_i}{3\rho_i}\sum_jm_j\frac{\partial W_{ij}(h_i)}{\partial h_i}\,,
\end{equation}

\noindent that will be used in the time derivatives, as well as the kernel imbalance according to Equation~\eqref{eq:Kernel_imbalance} that is needed for the interface correction (see Section~\ref{sec:Interface_correction}). \texttt{pkdgrav3} also supports a variant of SPH in which, instead of estimating the density through neighbor summation as described above, the density is evolved directly using the mass conservation equation (Equation~\eqref{eq:Mass_conservation}). This formulation can offer advantages in certain situations, such as at material interfaces or free surfaces, but it is less commonly used because it does not provide exact mass conservation. In this case, the density pass is still performed to calculate $h_i$ and $\Omega_i$.

\subsubsection{Interface correction}\label{sec:Interface_correction}
At interfaces between different materials and at vacuum interfaces, the density estimate provided by Equation~\eqref{eq:Density_estimate} can be extremely bad, leading to spurious forces \citep{woolfsonPracticalSPHModels2007,reinhardtNumericalAspectsGiant2017,dengPrimordialEarthMantle2019,dengEnhancedMixingGiant2019,reinhardtBifurcationHistoryUranus2020,ruiz-bonillaDealingDensityDiscontinuities2022}. To improve the density estimate, we use the method introduced by \citet{ruiz-bonillaDealingDensityDiscontinuities2022}, which is in turn based on \citet{reinhardtBifurcationHistoryUranus2020}. During the density pass, a kernel imbalance value is calculated for each particle which is defined as

\begin{equation}\label{eq:Kernel_imbalance}
I_i=\alpha\frac{\left\vert\sum_j\kappa_{ij}\left(\vec{r}_j-\vec{r}_i\right)m_jW_{ij}(h_i)\right\vert}{h_i\sum_jm_jW_{ij}(h_i)}\,,
\end{equation}

\noindent where $\kappa_{ij}$ is 1 if particles $i$ and $j$ are the same material and $-1$ if they are of different material and $\alpha$ is a normalization parameter. In the subsequent interface correction pass, this kernel imbalance $I_i$ is then used to calculate estimated average values for the temperature and pressure of the particle according to

\begin{equation}
\overline{T}_i=\frac{\sum_jT_je^{-I_j^2}W_{ij}(h_i)}{\sum_je^{-I_j^2}W_{ij}(h_i)}\,,\quad\overline{P}_i=\frac{\sum_jP_je^{-I_j^2}W_{ij}(h_i)}{\sum_je^{-I_j^2}W_{ij}(h_i)}\,.
\end{equation}

These average values are then used to calculate corrected temperature and pressure values

\begin{align}
\tilde{P}_i&=e^{-I_i^2}P_i+\left(1-e^{-I_i^2}\right)\overline{P}_i\,,\\
\tilde{T}_i&=e^{-I_i^2}T_i+\left(1-e^{-I_i^2}\right)\overline{T}_i\,,
\end{align}

\noindent which are then finally used to calculate a corrected density using the EOS

\begin{equation}
\tilde{\rho}_i = \rho_{EOS}(\tilde{T}_i,\tilde{P}_i)\,.
\end{equation}

Since there are cases where the interface correction is not needed (e.g., if there are no density discontinuities), the user can specify if it is active or not. If \texttt{pkdgrav3} is used with evolved instead of smoothed density (see Section~\ref{Density_and_smoothing_length_calculation}), interface correction can not be activated, as it specifically addresses a limitation of density smoothing and is therefore not applicable in this mode.

\subsubsection{Time derivatives}\label{sec:Time_derivatives}
The SPH representation of the momentum conservation equation (Equation~\eqref{eq:Momentum_conservation}) that results by derivation from the Lagrangian (see \citet{priceSmoothedParticleHydrodynamics2012} for a derivation) takes the form

\begin{equation}
\left.\frac{\upd \vec{v}_i}{\upd t}\right|_{H} =-\sum_jm_j\left[\frac{P_i}{\Omega_i\rho_i^2}\frac{\partial W_{ij}(h_i)}{\partial \vec{r}_i}+\frac{P_j}{\Omega_j\rho_j^2}\frac{\partial W_{ij}(h_j)}{\partial\vec{r}_i}\right]\,,
\end{equation}

\noindent where the subscript $H$ designates the contribution to the acceleration from the hydrodynamics. This expression contains both a gather contribution $W_{ij}(h_i)$ as well as a scatter contribution $W_{ij}(h_j)$ which necessitates to also add particles to the interaction list that may not be inside of the kernels of the particle in the current group but may have them inside of their kernel (see Section~\ref{sec:Neighbor_search}). In addition to the values of particle $i$, it also contains the density $\rho_j$, the pressure $P_j$ and the $\nabla h$ term $\Omega_j$ of the interaction particles. Thus, it has to be calculated in a separate pass, after all particles have their respective values updated.

One possible discretization of the energy conservation equation (Equation~\eqref{eq:Energy_conservation}) takes the form of this internal energy derivative

\begin{equation}
\left.\frac{\upd u_i}{\upd t}\right|_{H} =\frac{P_i}{\Omega_i\rho_i^2}\sum_jm_j(\vec{v}_i-\vec{v}_j)\cdot\frac{\partial W_{ij}(h_i)}{\partial\vec{r}_i}\,,
\end{equation}

\noindent but it is usually discarded in favor of strictly conserving entropy in the absence of shocks. If the built-in ideal gas implementation is used, this is done using the formulation described in \citet{springelCosmologicalSmoothedParticle2002}. For all other EOSs that support this, the formulation described in \citet{reinhardtNumericalAspectsGiant2017} is used. This scheme uses the old density and internal energy to calculate the isentropic evolution to the new density, by having the EOS solve $u_{new} = u(\rho_{new},s(\rho_{old},u_{old}))$, before applying the closing kick (see Section~\ref{sec:Time_Integration}).

In order to capture shocks, we use artificial viscosity in the original SPH form \citep{monaghanSmoothedParticleHydrodynamics1992}

\begin{equation}
\Pi_{ij} = \begin{cases}\frac{-\alpha \overline{c}_{ij}\mu_{ij}+\beta\mu_{ij}^2}{\overline{\rho}_{ij}}&\mbox{for } \vec{v}_{ij}\cdot\vec{r}_{ij}<0\\
0&\mbox{otherwise}\end{cases}\,,
\end{equation}

\noindent where

\begin{equation}
\mu_{ij}=\frac{\overline{h}_{ij}(\vec{v}_{ij}\cdot\vec{r}_{ij})}{\vec{r}_{ij}^2+\epsilon \overline{h}_{ij}^2}\,,
\end{equation}

\noindent where $\overline{c}_{ij}$, $\overline{\rho}_{ij}$ and $\overline{h}_{ij}$ are the averages of the respective values for particles $i$ and $j$. $\alpha$ and $\beta$ represent the shear and Von Neumann-Richtmyer viscosities respectively. The artificial viscosity generates contributions (denoted by the subscript $\Pi$) to both the accelerations

\begin{equation}
\left.\frac{\upd\vec{v}_i}{\upd t}\right|_{\Pi} = -\sum_jm_j\Pi_{ij}\frac{\partial \overline{W}_{ij}}{\partial\vec{r}_i}\,,
\end{equation}

\noindent and the internal energy derivatives

\begin{equation}\label{eq:AV_heating}
\left.\frac{\upd u_i}{\upd t}\right|_{\Pi}  = \frac{1}{2}\sum_jm_j\Pi_{ij}(\vec{v}_i-\vec{v}_j)\cdot\frac{\partial \overline{W}_{ij}}{\partial\vec{r}_i}\,,
\end{equation}

\noindent where we use the averaged kernel $\overline{W}_{ij} = 0.5(W_{ij}(h_i) + W_{ij}(h_j))$. This contribution to the internal energy derivative is active independent if the entropy conservation scheme is active or not. To reduce the effect of the artificial viscosity outside of shocks different viscosity switches have been developed \citep[e.g.,][]{balsaraNeumannStabilityAnalysis1995,morrisSwitchReduceSPH1997,cullenInviscidSmoothedParticle2010} that all have their mix of advantages and disadvantages. As we show in Sections~\ref{sec:Gresho-Chan_vortex} and~\ref{sec:Kelvin-Helmholtz_instability} having the full artificial viscosity active in non-shocking environments has profound effects on the simulation results. Some of these effect can be compensated by higher resolution, as the effective viscosity caused by artificial viscosity depends on the resolution.

If gravity is active, the body force term in Equation~\eqref{eq:Momentum_conservation} gives an additional contribution to the accelerations $\left.\frac{\upd \vec{v}_i}{\upd t}\right|_{G}$ which is calculated with FMM (see Section~\ref{sec:Gravity}) in the same pass, using the same particle-particle interaction list. This slightly increases the work for gravity compared to a pure gravity FMM pass without SPH, as the PP interaction list is larger, but this increase is much smaller than doing an additional pass just for gravity.

\subsubsection{Time integration}\label{sec:Time_Integration}
\begin{figure}[ht!]
\centering
\includegraphics[width=\linewidth]{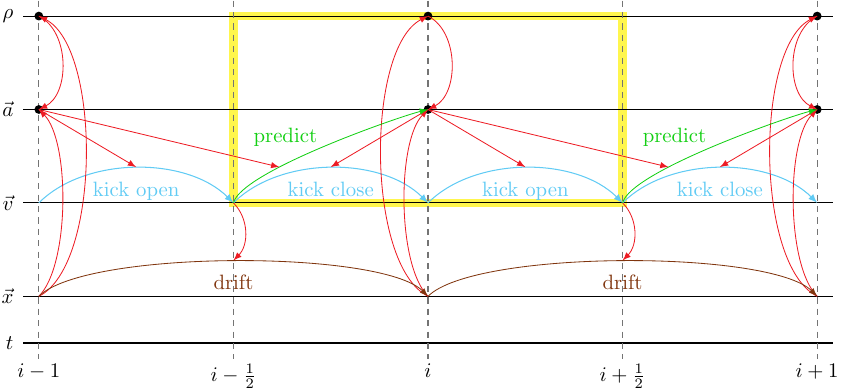}
\caption{Illustration of how a single time step is done (the kick phase is indicated by the yellow box). First, the density update is done for all particles (or the active particles and their neighbors in the case of FastGas, see the text). Then, using the time derivatives from the last step, the velocity and internal energy are predicted (green arrows). Using these values, the force calculation, the closing kick of the last step and the opening kick of the new step are done in succession (light blue arrows), asynchronously for each group of active particles. After these updates are done, the positions are drifted using the new velocities (brown arrows). The red arrows imply a flow of information.}
\label{fig:Time_Integration}
\end{figure}

We use a hierarchical kick-drift-kick leap-frog scheme to integrate the velocities and positions in time. Depending on the time stepping criteria, each particle is integrated with a time step size that is a power-of-two fraction of the global timestep $\Delta t_0$ (this is called a substep, with the rung value giving the power of two). While the leap-frog scheme is well suited if the force does not depend on the velocity (as is the case for gravity), hydrodynamic forces depend on the velocity and the internal energy which are not known at the time they are needed. To solve this problem, we embed a predictor-corrector scheme into the leapfrog scheme by predicting the velocity and internal energy from the last step, using the time derivatives from the last step, to calculate the new time derivatives. The sequence of operations for a single substep is illustrated in Figure~\ref{fig:Time_Integration} and consists of the steps

\begin{enumerate}
    \item Density pass: Calculate the density for all particles at time $i$.
    \item Predict the velocity and internal energy from time $i-\frac{1}{2}$ to time $i$ using the derivatives from time $i-1$
    \begin{align}
        \vec{v}_{i,pred} &= \vec{v}_{i-\frac{1}{2}}+\frac{1}{2}\Delta t \vec{a}_{i-1}\,,\\
        u_{i,pred} &= u_{i-\frac{1}{2}}+\frac{1}{2}\Delta t \dot{u}_{i-1}\,,
    \end{align}
    \noindent and calculate fluid values ($P$, $T$, $c$).
    \item Density correction pass: Perform the interface correction for all particles if it is activated, update fluid values ($P$, $T$, $c$).
    \item Forces pass: For all particles with rung larger than or equal to the rung of the substep \begin{itemize}
        \item Calculate the time derivatives at time $i$ using the predicted values.
        \item Perform the closing kick from time $i-\frac{1}{2}$ to time $i$ for the velocities and the internal energy
        \begin{align}
        \vec{v}_{i} &= \vec{v}_{i-\frac{1}{2}}+\frac{1}{2}\Delta t \vec{a}_{i}\,,\\
        u_{i} &= u_{i-\frac{1}{2}}+\frac{1}{2}\Delta t\dot{u}_{i}\,.
        \end{align}
        \item If necessary and possible, change the time step rung of the particle based on the time step criterion.
        \item Perform the opening kick from time $i$ to $i+\frac{1}{2}$ for the velocities and the internal energy
        \begin{align}
        \vec{v}_{i+\frac{1}{2}} &= \vec{v}_{i}+\frac{1}{2}\Delta t \vec{a}_{i}\,,\\
        u_{i+\frac{1}{2}} &= u_{i}+\frac{1}{2}\Delta t \dot{u}_{i}\,.
        \end{align}
    \end{itemize}
    \item Drift the positions of all particles from time $i$ to time $i+1$
    \begin{equation}
        \vec{x}_{i+1} = \vec{x}_{i}+\Delta t \vec{v}_{i+\frac{1}{2}}\,.
    \end{equation}
\end{enumerate}

Between the closing kick and the opening kick, the rung of a particle can change, based on its maximum allowed time step size. This value is calculated using two criteria. The first is based on the acceleration: $\Delta t_{i,a}=\eta_a\sqrt{\epsilon/\left\vert a_i\right\vert}$ where $\eta_a=0.2$ and $\epsilon$ is either the gravitational softening length if gravity is active, or the SPH kernel size. To ensure that the Courant-Friedrichs-Lewy (CFL) condition is satisfied, the second criterion is

\begin{equation}\label{eq:courant_condition}
\Delta t_{i,CFL}=\eta_C\min_j\frac{h_i}{(1+0.6\alpha)c_i+0.6\beta\left\vert\min(0,\mu_{ij})\right\vert}\,,
\end{equation}

\noindent where $\eta_C$ is the Courant parameter and $\alpha$ and $\beta$ are the artificial viscosity parameters \citep{monaghanSmoothedParticleHydrodynamics1992,wadsleyGasoline2ModernSmoothed2017}. This maximum allowed time step size then gives a desired rung $r_i$ for the particle such that

\begin{equation}
\min\left(\Delta t_{i,a}, \Delta t_{i,CFL}\right)\geq 2^{-r_i}\Delta t_0\,.
\end{equation}

\noindent Additionally, we enforce that no particle has a time step size that is larger than four times that of all particles it interacts with, by increasing the rungs such that $r_i\geq (r_j-2)$ for all neighbors $j$. This condition was first introduced by \citet{saitohNECESSARYCONDITIONINDIVIDUAL2009} to improve energy conservation. This usually ensures stable integration, except for pathological cases, where particle on very different rungs are interacting, for example the Sedov-Taylor blast wave test (see Section~\ref{Sedov-Taylor_blast_wave} for a discussion). A particle can always move to a higher rung with a smaller time step, but moving to a lower rung is only possible on substeps that correspond to the rung it wants to move to. This means that if the minimum time step size of a particle increases, it may have to wait a few substeps before it can move to the new step size.

This hierarchical ladder of time step rungs can lead to situations where the highest rungs (with the smallest time step sizes) are only populated by a small fraction of the total particles. Instead of updating the density for all particles, we employ a procedure we call FastGas if the fraction of active particles is below a threshold. Before the density and density correction passes we perform a marking pass, where we mark all particles that would end up on interaction lists of active particles to find particles that need an updated density. If density correction is active, we perform a second marking pass where we mark all particles that would end up on interaction lists of these marked particles to find those that need an initial density estimate. Then, the density pass is performed only for those particles with the first mark (or either mark if density correction is active), and the density correction pass is performed only for particles with the first mark. This significantly speeds up substeps with low rung population, but it only makes sense if the population is really small, as the marking passes are not free, because they also need to walk the tree.

\subsubsection{File format and Initial conditions}
Currently, \texttt{pkdgrav3} uses the \texttt{TIPSY}\footnote{\url{https://faculty.washington.edu/trq/hpcc/tools/tipsy/tipsy.html}} format to read initial conditions and to store simulation outputs, which are written at user-defined intervals. Each output file contains the particle mass $m$, position $\vec{x}$, velocity $\vec{v}$, density $\rho$, temperature $T$, smoothing length $h$, material ID, and gravitational potential $\phi$, corresponding to 48 bytes per particle. At user-defined intervals, and at the end of a queue shot on queue-based systems, a checkpoint is written as a dump of each thread's particle storage, allowing the simulation to restart seamlessly. These checkpoint files include all quantities stored by the code and require 120 bytes per particle. In addition to the values contained in the \texttt{TIPSY} files, they store the acceleration $\vec{a}$, the $\nabla h$ term $\Omega$, the velocity divergence $\nabla\cdot\vec{v}$, the internal energy $u$, its time derivative $\frac{\upd u}{\upd t}$, the previous density used in the entropy-conserving scheme $\rho_{old}$, the interface correction imbalance in the form $e^{-I^2}$, the maximum group size used by the dynamic group size scheme, as well as the predicted sound speed $c_{pred}$, pressure $P_{pred}$, temperature $T_{pred}$, and velocity $\vec{v}_{pred}$.

The \texttt{TIPSY} file containing the initial conditions can be generated as required for the problem under study. The examples shown in Section~\ref{sec:Hydrodynamic_tests} are created using \texttt{numpy}, while the planetary models used in Section~\ref{sec:Scaling_tests} are generated with \texttt{ballic} \citep{reinhardtNumericalAspectsGiant2017,chauFormingMercuryGiant2018,reinhardtBifurcationHistoryUranus2020}. The \texttt{ballic} code solves the equations of hydrostatic equilibrium to obtain 1D density and temperature profiles. It then creates a particle representation of these profiles by distributing spherical \texttt{HEALPix} \citep{gorskiHEALPixFrameworkHighResolution2005} shells of equal-mass particles and optimizing the radial to tangential axis ratios of the cells containing the individual particles. As a result, the smoothed density is in excellent agreement with the 1D density profile. Finally, each particles temperature is determined from the 1D equilibrium model.

The particle realizations produced by \texttt{ballic} exhibit very low velocity noise \citep{reinhardtNumericalAspectsGiant2017}; however, at high resolutions some relaxation is often still required. To facilitate this, \texttt{pkdgrav3} provides an option to damp particle velocities before applying the opening kick and after prediction. The damping follows an exponential form,

\begin{equation}
\vec{v} \longrightarrow \vec{v}e^{-\frac{\Delta t}{\Delta t_0}f_{VD}}\,,
\end{equation}

\noindent ensuring that the cumulative effect of the damper is consistent across particles on different time step rungs. The damping factor $f_{VD}$ may be held constant or ramped up or down over the course of the simulation. When the velocity damper is active, shock heating is disabled (Equation~\eqref{eq:AV_heating} is set to zero), allowing fully isentropic relaxation. Because the velocity damper breaks conservation of energy, linear momentum, and angular momentum, it should only be used for preparing relaxed initial conditions.

Since \texttt{ballic} produces spherically symmetric bodies, \texttt{pkdgrav3} also includes a tool to transform these into rotationally flattened configurations. This is achieved by evolving the body in the co-rotating reference frame following the approach of \citet{timpeMachineLearningApplied2020}. In this frame, rotation manifests as a centrifugal acceleration which is applied during the force pass, before applying the kicks to the velocities. The target angular velocity is gradually increased to its final value to minimize overshoot, and the process can be further stabilized by combining it with the velocity damper.

\section{Hydrodynamic tests}\label{sec:Hydrodynamic_tests}
In this section, we show how our code performs in several commonly used benchmarks, to demonstrate that it behaves as expected from a traditional SPH code. Being a traditional SPH implementation, the code suffers from the typical shortcomings of the SPH method, but the increased resolution our code is capable of is able to reduce some of them. All initial conditions, with the exception of the Evrard collapse test (see Section~\ref{sec:Evrard_collapse}) which uses a glass-type initial condition, are built based on a grid of body centered cubic cells (BCC), but we compare the soundwaves test to the results of a primitive cubic grid. Usually, SPH prefers glass-type initial conditions, especially when using low neighbor counts, but because we mostly use the Wendland C6 kernel \citep{dehnenImprovingConvergenceSmoothed2012} with 400 neighbors, this point is not critical for the tests. Unlike for grid codes, tests in one or two dimensions do not accurately test the behavior of SPH in a three-dimensional simulation. Thus, in tests that are inherently of lower dimension, we use pseudo-1D or pseudo-2D setups, where the reduced dimensions are represented by a small number of cells in that direction, between two periodic boundaries. The small number of cells in reduced-dimension directions is chosen to be 16 for the lowest density region of the test domain, in order to ensure that under no circumstances any of the periodic replicas of a particle can show up in its own kernel, as this would introduce a strong coupling that could lead to artifacts. All tests were performed in a unit system where $[L] = \SI{1}{\Rearth}$, $[V]=\SI{1}{\kilo\meter\per\second}$ and $[G]=1$.

\subsection{Linear traveling sound wave}
\begin{figure}[ht!]
\centering
\includegraphics[width=\linewidth]{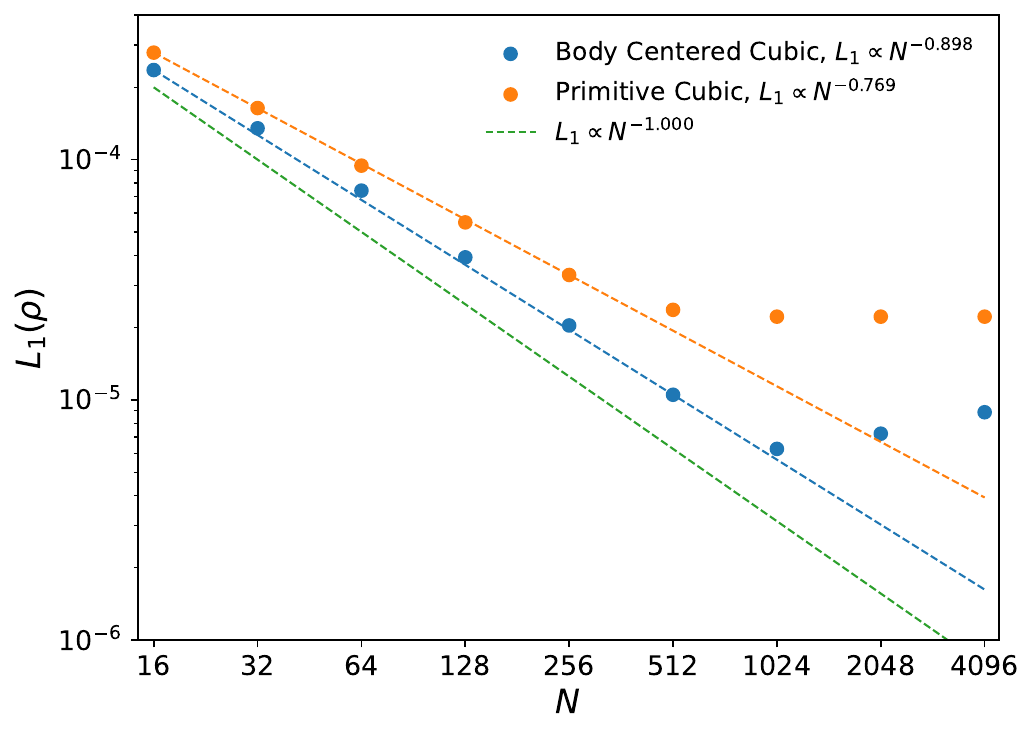}
\caption{$L_1$ error norm of the linear traveling sound wave problem. In blue, the body centered cubic grid shows $N^{-0.898}$ scaling with the number of grid cells in the direction of wave propagation, while in orange the primitive cubic grid shows $N^{-0.769}$ scaling. As expected for traditional SPH, the scaling falls short of first-order and becomes dominated by the floating-point precision used by the code above $N=1024$ (see text for discussion).}
\label{fig:sound_waves}
\end{figure}

We start with a simple linear one-dimensional sound wave. Following \citet{stoneAthenaNewCode2008}, we create a periodic domain of unit length with a polytropic $\gamma=5/3$ gas with unit density and sound speed, such that the background pressure is $P=3/5$. We sample the domain with $N$ BCC grid cells in the $x$-direction and 16 cells in $y$- and $z$-direction, the reduced dimensions mentioned previously. We then add a traveling sound wave with a small amplitude of $A = \delta \rho / \rho = 10^{-3}$ with unit wavelength. The density wave is imprinted by slightly moving each particle according to $\Delta x = -\frac{A}{2\pi}\cos(2\pi x - \pi / 2)$. Because the code currently stores all values, except for positions, and executes all computations in single precision, we need to choose a larger amplitude $A$ than other publications \citep[e.g.,][]{stoneAthenaNewCode2008,hopkinsNewClassAccurate2015} to ensure that the wave is sampled with enough accuracy to not introduce dominating round-off errors (see also the discussion in \citet{alonsoasensioMeshfreeHydrodynamicsPkdgrav32023} where the authors have the same problem). The pressure and temperature values are then set according to $P=P_0(1+A\gamma\cos(2\pi x))$ and $T=\frac{P}{\rho R_{gas}}$. Figure~\ref{fig:sound_waves} shows the $L_1$ error norm of the resulting density wave after it passed the domain once for different values of $N$. The error decreases with $N^{-0.898}$ if a BCC grid is used and with $N^{-0.769}$ for a primitive cubic grid, consistent with the expected convergence of standard SPH (i.e. close to first order, \citet{dehnenImprovingConvergenceSmoothed2012, zhuNUMERICALCONVERGENCESMOOTHED2015}). In the BCC case, the error follows this scaling until $N=1024$, where the errors produced by the floating-point precision of the code start to dominate and distort the solution.

\subsection{Sod shock tube}
\begin{figure}[ht!]
\centering
\includegraphics[width=\linewidth]{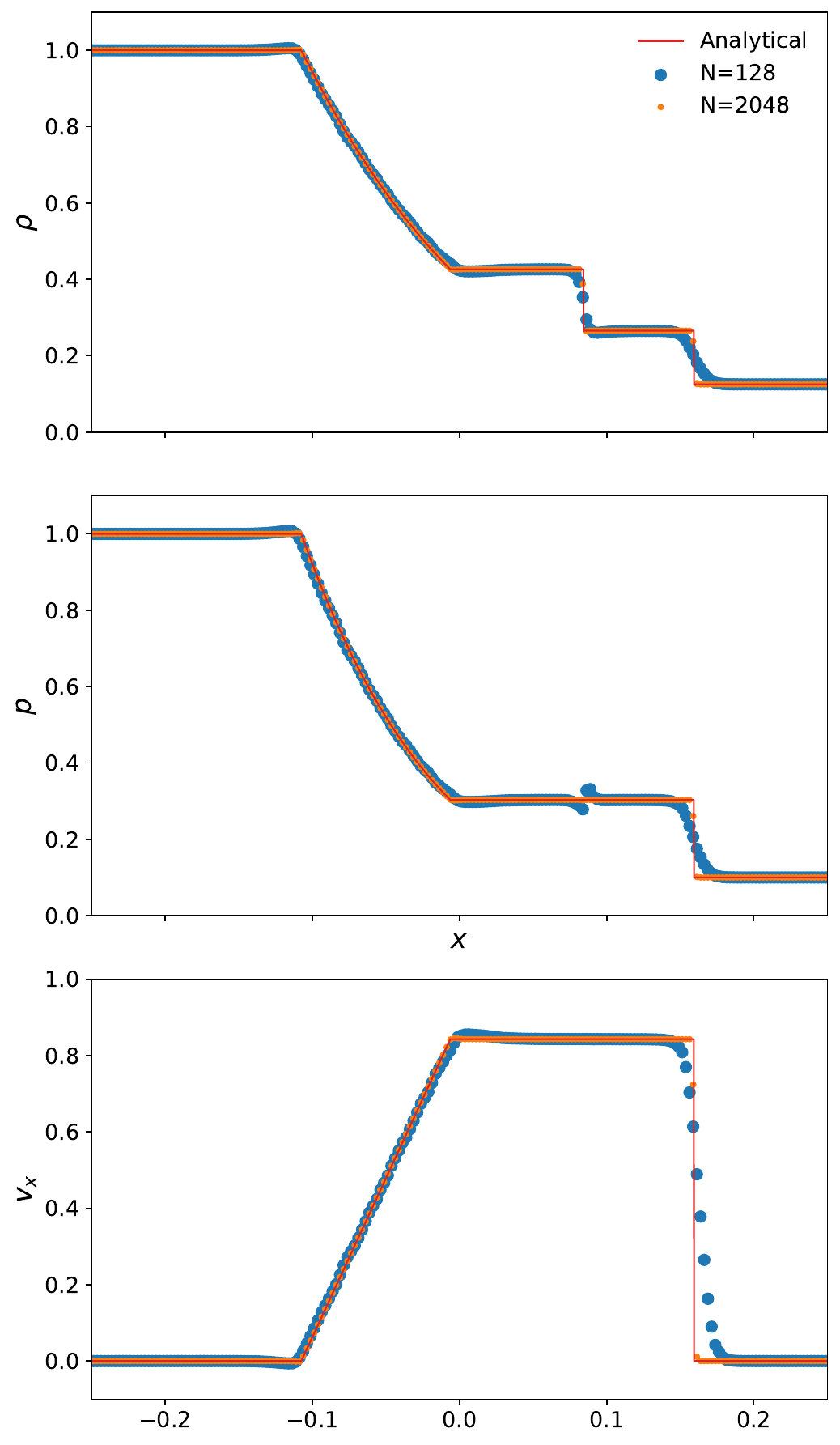}
\caption{Results of the Sod shock tube test for two different numbers $N$ of particles sampling the $x$-direction of the low-density region. The values are binned in 250 bins. While the low-resolution ($N=128$) solution still shows the typical SPH features (smoothed discontinuities, pressure blip), the high-resolution ($N=2048$) solution follows the analytical solution closely and the pressure blip is all but gone.}
\label{fig:sod_shock_tube}
\end{figure}

The classic Sod shock tube \citep{sodSurveySeveralFinite1978} is a standard test employed by many authors to show how the code handles the shock evolution in this simple 1D Riemann problem. The initial condition is defined as a $\gamma=7/5$ gas at rest with two distinct regions in a periodic domain of unit length. In the left region ($x<0$), the density and pressure are $\rho_L=P_L=1.0$, while in the right region ($x\geq0$), the density is $\rho_R=0.125$ and the pressure is $P_R=0.1$. This leads to a contact discontinuity and a shockwave traveling to the right as well as a rarefaction wave traveling to the left. We sample the low-density (right) region with an $N\times 16^2$ BCC grid. The high-density (left) region is sampled with $2N\times 32^2$ cells to achieve a density contrast of $8$. Figure~\ref{fig:sod_shock_tube} shows the density, velocity and pressure results for two different values of $N$ and compares them to the analytical solution. In both cases, the Wendland C6 kernel with 400 neighbors is used, and the artificial viscosity parameters are set to $\alpha=1.5$ and $\beta=3.0$. While the lower resolution result ($N=128$) shows visible smoothing of the steps, the high-resolution ($N=2048$) solution follows the analytical solution nearly perfectly. The pressure blip at the contact discontinuity, a typical feature of SPH, that is clearly visible at $N=128$ all but vanishes at $N=2048$.

\subsection{Sedov-Taylor blast wave}\label{Sedov-Taylor_blast_wave}
\begin{figure}[ht!]
\centering
\includegraphics[width=\linewidth]{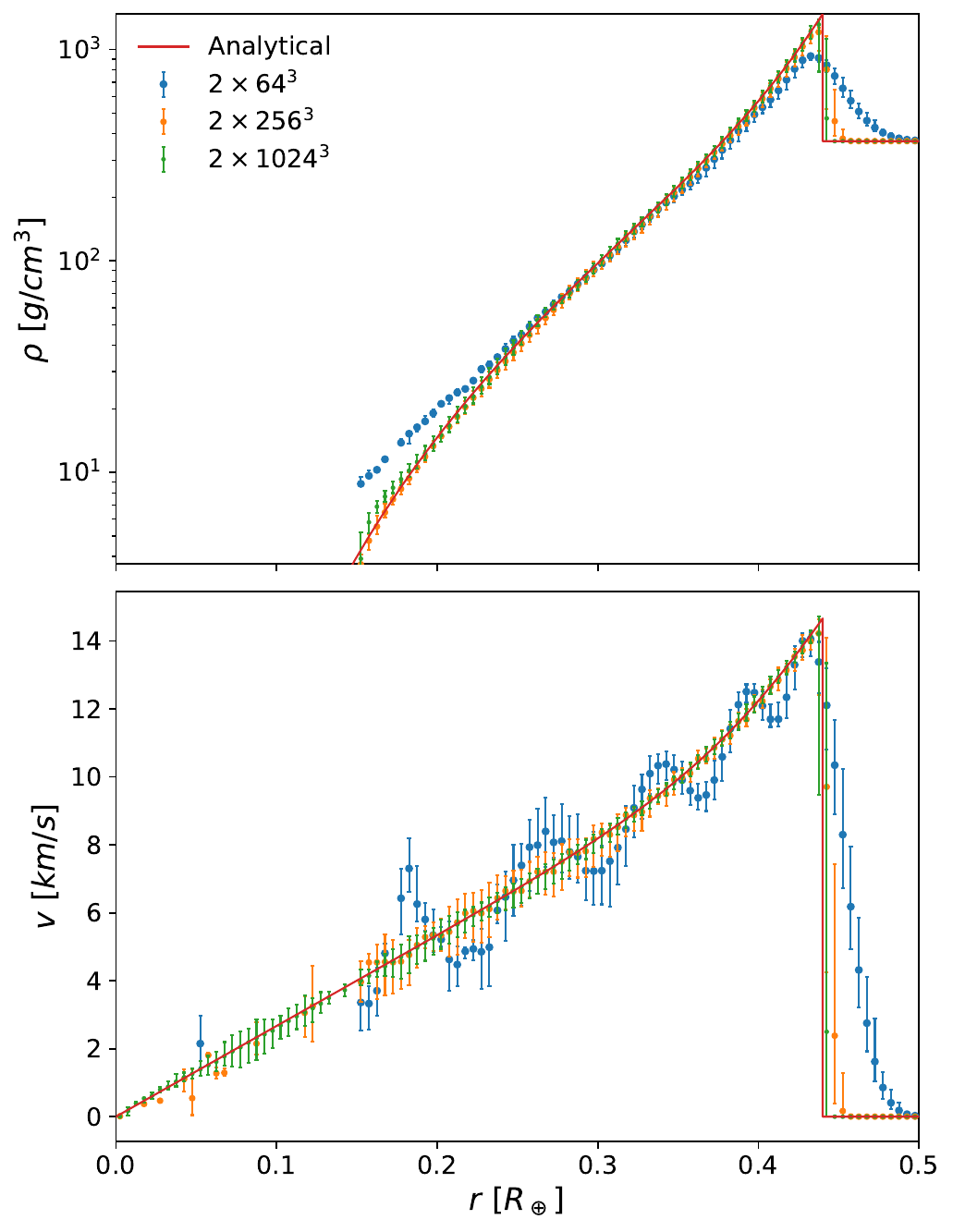}
\caption{Results of the Sedov-Taylor blast wave test for different resolutions. The top panel shows the density while the bottom panel shows the radial velocity. All values are binned in 100 bins, with the marker showing the average value in the bin while the error bars show maximum and minimum values.}
\label{fig:sedov_taylor_blastwave_errorbars}
\end{figure}

The Sedov-Taylor blast wave \citep{sedovPropagationStrongShock1946,taylorFormationBlastWave1950,taylorFormationBlastWave1950a,sedovSimilarityDimensionalMethods1959} is a spherical shock wave generated by the injection of energy in a small central region. For this test, we generate a BCC grid in a $1\times 1\times 1$ box with periodic boundary conditions. The gas has a density of $\rho=1$ and the temperature of the particles in a central sphere of radius $r=0.02$ is set such that the total injected energy is \SI{9.56072e23}{\joule}, while the surrounding gas has a temperature of \SI{10}{\kelvin}. This setup results in an infinite Mach number shock with a density enhancement of 4 for $\gamma=5/3$. Using a time stepping scheme with individualized time step sizes per particle leads to large errors in energy conservation as demonstrated by \citet{saitohNECESSARYCONDITIONINDIVIDUAL2009}, because particles with widely different time step sizes interact with each other. The solution proposed by \citet{saitohNECESSARYCONDITIONINDIVIDUAL2009} and described in Section~\ref{sec:Time_Integration}, does not work in this situation as particles on the largest timestep should interact with the expanding shockwave at times where they do not get updates. We thus chose to use a dynamic global time step size such that it is always smaller than the smallest needed by any particle at any given time, to circumvent this problem. We like to note that this setup is artificial and in real simulations, such large discontinuities in temperature and pressure will not occur. With this setting, the total energy conservation error over the full simulation is below \SI{0.2}{\percent} even at the highest resolution, but this value increases with resolution. Figure~\ref{fig:sedov_taylor_blastwave_errorbars} shows the results of the Sedov-Taylor blast wave test for three different resolutions. The result of the lowest resolution ($2\times64^3$ particles, blue markers) clearly shows that the shock is strongly smoothed by the large kernel used and it also exhibits strong post-shock ringing \citep{huSPHGalSmoothedParticle2014} in the velocity. These effects are significantly reduced at higher resolution and the solution at $2\times1024^3$ follows the analytical solution nearly perfectly.

\subsection{Evrard collapse}\label{sec:Evrard_collapse}
\begin{figure}[ht!]
\centering
\includegraphics[width=\linewidth]{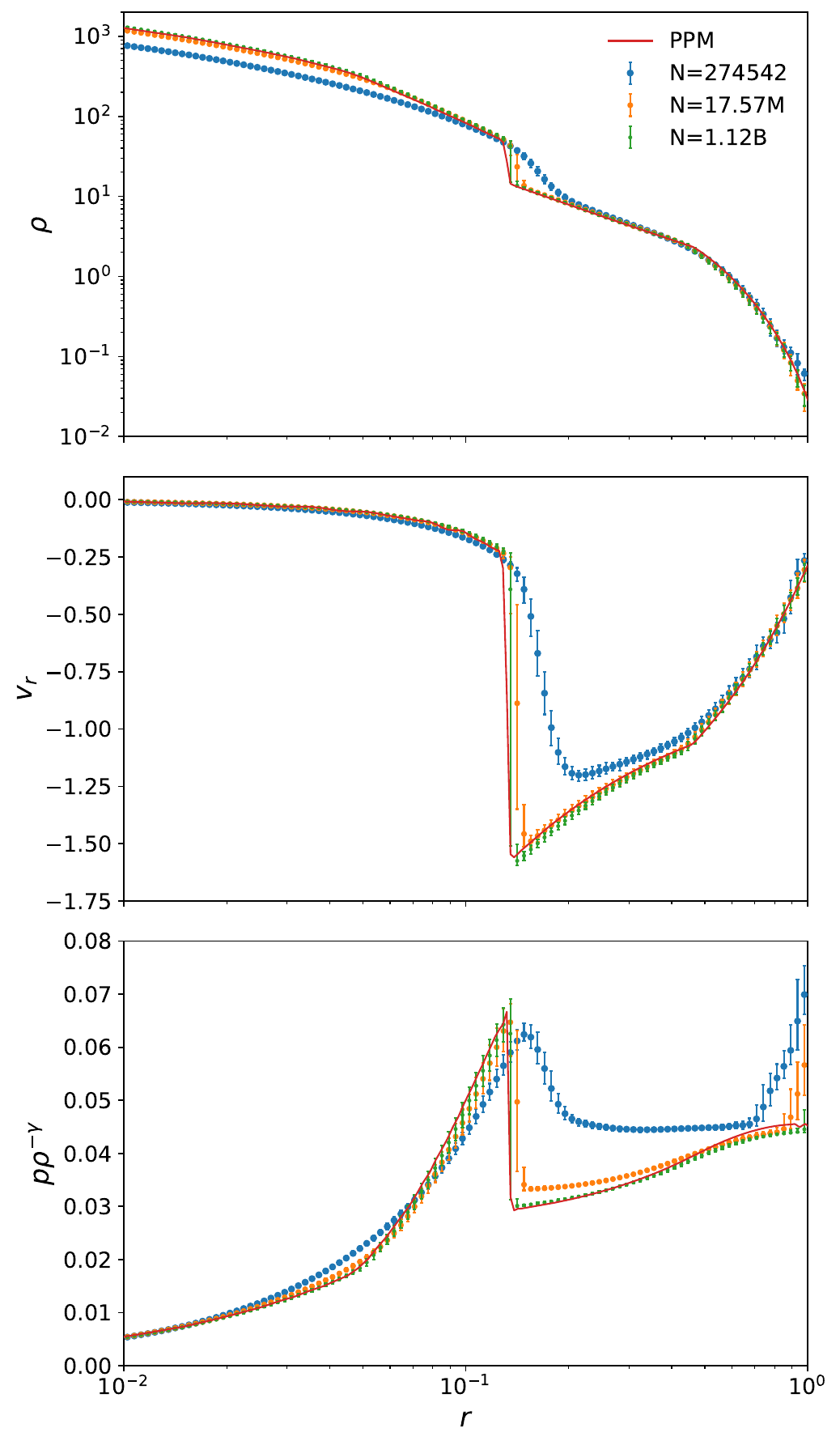}
\caption{Results of the Evrard collapse test for different resolutions. The top panel shows the density, the middle panel the radial velocity and the bottom panel the entropy. All values are binned in 100 bins, with the marker showing the average value in the bin while the error bars show maximum and minimum values.}
\label{fig:evrard_profiles_errorbars}
\end{figure}

\begin{figure}[ht!]
\centering
\includegraphics[width=\linewidth]{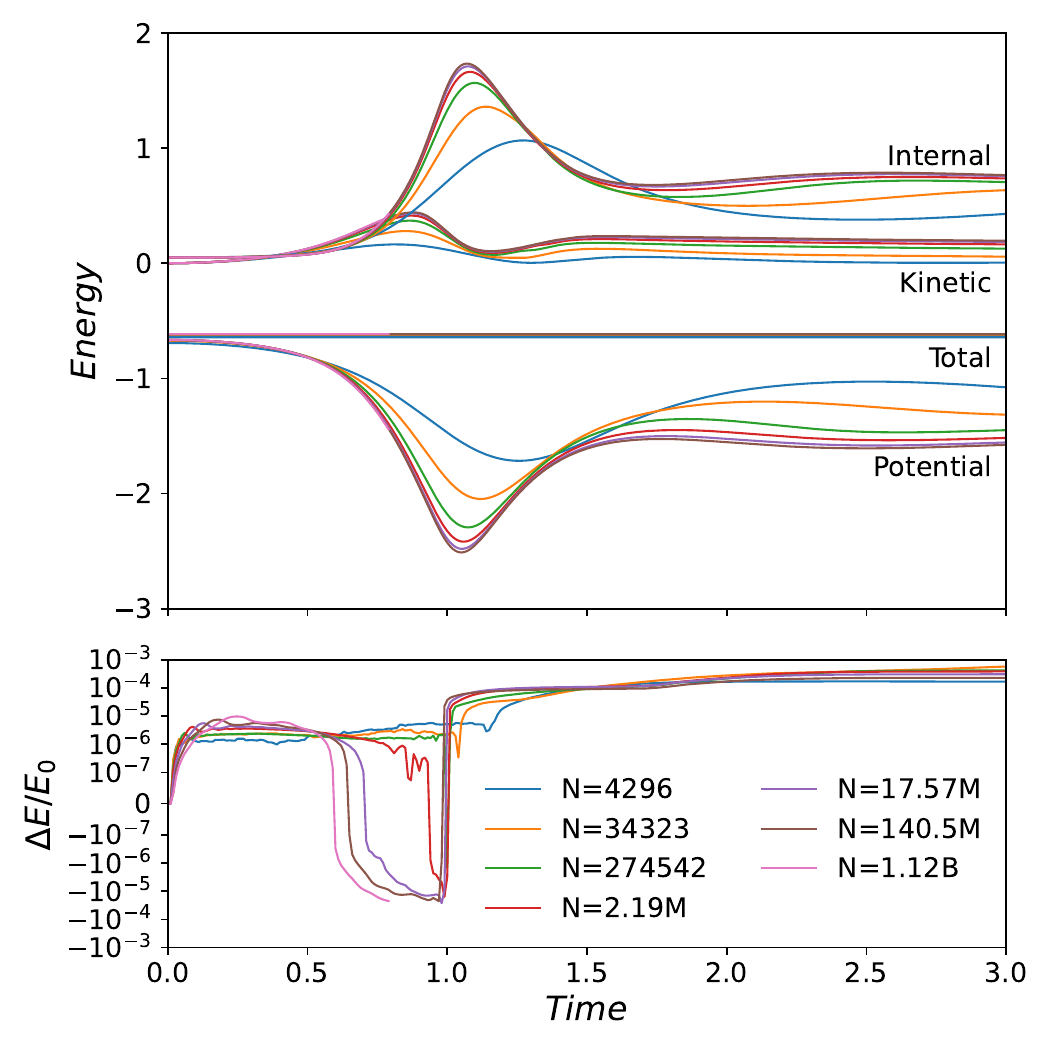}
\caption{Top panel: time evolution of the potential, kinetic, internal and total energy in Evrard's spherical collapse test for different resolutions. The simulation with $N=1.12$ billion particles is only run up to $t=0.8$. Bottom: The total energy is conserved to within \SI{0.06}{\percent}, even during shock formation, with conservation improving at higher resolution.}
\label{fig:evrard_energies}
\end{figure}

To test the coupling to the gravity solver of \texttt{pkdgrav3} we use the adiabatic collapse test from \citet{evrardNbody3DCosmological1988}. It models the cold infall facilitated by the self-gravity of an out-of-equilibrium gas cloud followed by a strong outward shock. The initial conditions consist of a sphere with density $\rho(r)=M/(2\pi R^2r)$ at rest, with an initial energy per unit mass of $0.05$. The sphere has radius $R=1$ with an enclosed mass of $M=1$. To construct the initial particle arrangement, we first create a $64^3$ BCC lattice which is then evolved into a glass-like arrangement as this test is extremely sensitive to any structure in the initial particle distribution that gets amplified by the slow infall. This cube is then tiled as necessary and a sphere is cut. The particle radii are then rescaled by $r\rightarrow r^{0.5}$ to produce the $1/r$ density profile. Figure~\ref{fig:evrard_profiles_errorbars} shows the solutions at $t=0.8$ for different resolutions together with the solution of a piecewise parabolic (PPM) 1D code by \citet{steinmetzCapabilitiesLimitsSmoothed1993}. At a resolution of $N=274542$ particles, the solution shows large deviations from the PPM solution, with the central density being underestimated, the shock position being too far out and the unshocked material outside of the shock shows increased entropy production as it is heated by the artificial viscosity triggered by the compression \citep{wadsleyGasoline2ModernSmoothed2017}. At higher resolutions, these effects are strongly reduced, with the solution at $N=1.12$ billion particles following the PPM solution closely. Figure~\ref{fig:evrard_energies} shows the evolution of the energy distribution between gravitational, thermal and kinetic energy during the simulation with the energy curves slowly converging with increasing resolution. It also shows the energy conservation error over the simulation. In order to eliminate differences in the energy evolution arising from resolution-dependent timestep sizes, we adopt a global timestep that is smaller than the minimum required by any particle, analogous to the Sedov-Taylor blast wave test. This leads to energy conservation better than \SI{0.06}{\percent} at $N=34323$, but contrary to the Sedov-Taylor blast wave test, here the energy conservation error decreases for higher resolutions. When individual particle timesteps are used in the \SI{140.5e6}{} particle simulation, particles are allowed to take significantly larger steps. This substantially reduces the total runtime but increases the energy conservation error to approximately \SI{1}{\percent}. In giant-impact simulations, conservation errors remain consistently below \SI{1}{\percent}. The trade-off between computational cost and conservation accuracy can be tuned with the parameter $\eta_C$ (see Equation~\eqref{eq:courant_condition}).

\subsection{Gresho-Chan vortex}\label{sec:Gresho-Chan_vortex}
\begin{figure}[ht!]
\centering
\includegraphics[width=\linewidth]{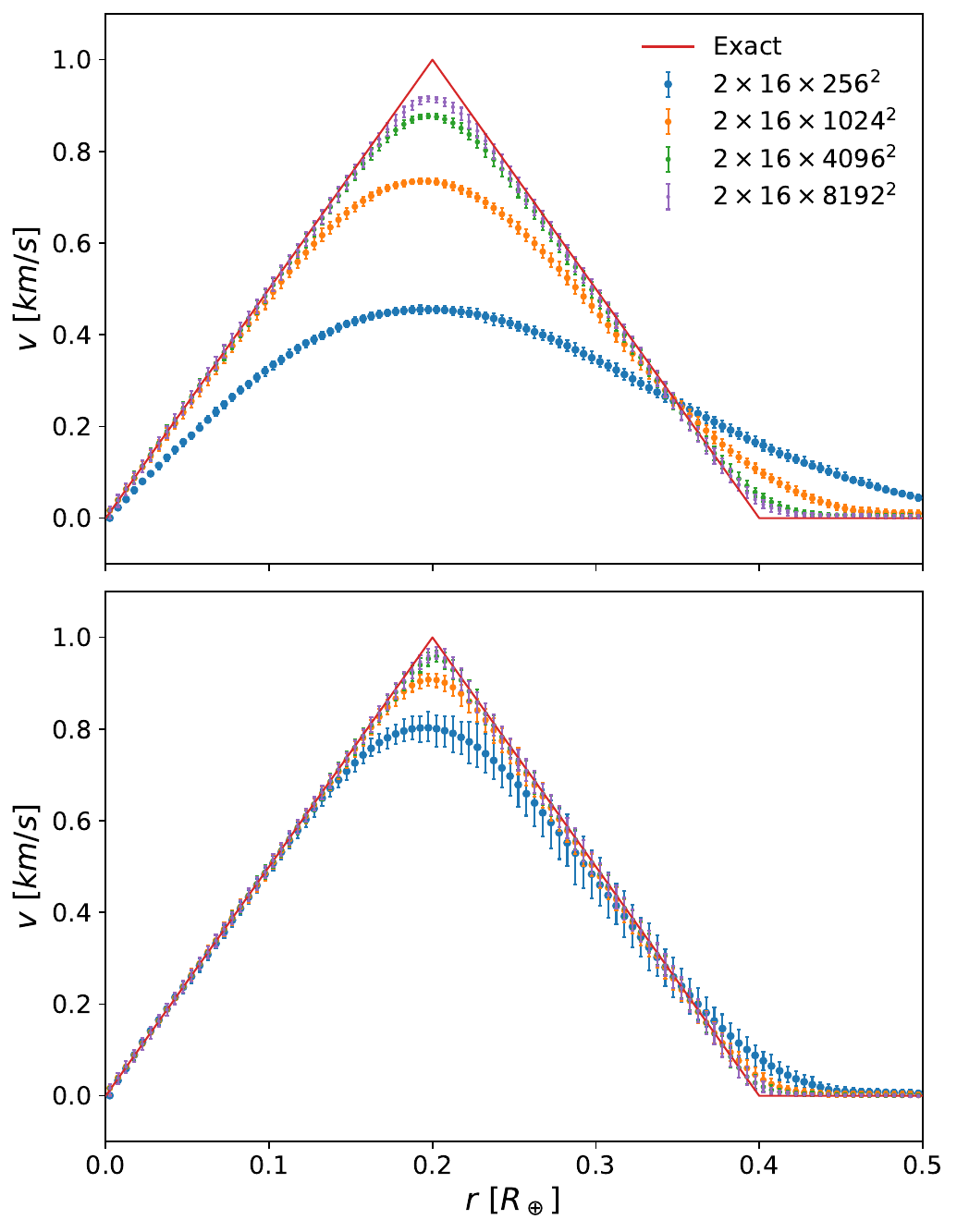}
\caption{Velocity profiles of the Gresho-Chan vortex at different resolutions at time $t=3.0$. The top panel shows the solutions with the viscosity parameters $\alpha=1.0$ and $\beta=2.0$ while the bottom panel shows the solutions with $\alpha=0.1$ and $\beta=0.0$. It is clear that the lower values give the better results, but in both cases increasing the resolution brings the solution closer to the exact solution. The velocity values are binned in 100 bins. The marker shows the average value and the error bars show maximum and minimum values in the bin. Higher viscosity values give solutions with less scatter.}
\label{fig:vortex_errorbars}
\end{figure}

The Gresho-Chan vortex \citep{greshoTheorySemiimplicitProjection1990} uses a cylindrical vortex to test how well a hydrocode can simulate an inviscid fluid. It has an exact solution that is steady as centrifugal accelerations are perfectly balanced by the pressure gradient. Artificial viscosity will cause angular momentum transport which disrupts this equilibrium, torquing up the outer parts of the vortex at the expense of the inner parts. The initial condition is defined in a periodic domain with $-1< x,y < 1$ which we sample with an $N^2\times 16$ BCC grid. The density is homogeneous with $\rho=1$ and the velocity is defined by the piecewise function:

\begin{equation}
v_\theta(r)=\begin{cases}5r&0\leq r<0.2\\2-5r&0.2\leq r < 0.4\\ r\geq 0.4\end{cases}
\end{equation}

\noindent which is stabilized by the pressure function:

\begin{equation}
P(r) = \begin{cases}5+12.5r^2&0\leq r< 0.2\\9+12.5r^2-20r+4\ln5r&0.2\leq r<0.4\\3+4\ln2&r\geq 0.4\end{cases}\,.
\end{equation}

We evolve the vortex to $t=3$, which corresponds to $\sim2.4$ rotations of the peak. As our code does not use a viscosity limiter, we provide the solutions with two different sets of viscosity parameters. The top panel of Figure~\ref{fig:vortex_errorbars} shows the solutions of the simulation with $\alpha=1.0$ and $\beta=2.0$ at four different resolutions, while the bottom panel shows the results at the same resolutions but with $\alpha=0.1$ and $\beta=0.0$. It is clear that the second set of viscosity parameters gives better results than the first one, but one can see that increased resolution brings the solution in both cases closer to the exact solution and higher viscosity values give solutions with less scatter. The average solution at $2\times 16\times8192^2$ with the large viscosity values is very similar to the average solution with the small viscosity values at a resolution of $2\times 16\times 1024^2$ (although it has smaller scatter), so it seems that the higher viscosity values needed to capture shocks can be ``compensated'' by increasing the linear resolution by a factor of 8 (or 512 times as many particles in a full 3D simulation).

\subsection{Hydrostatic equilibrium test}\label{sec:Hydrostatic_equilibrium_test}
\begin{figure}[ht!]
\centering
\includegraphics[width=\linewidth]{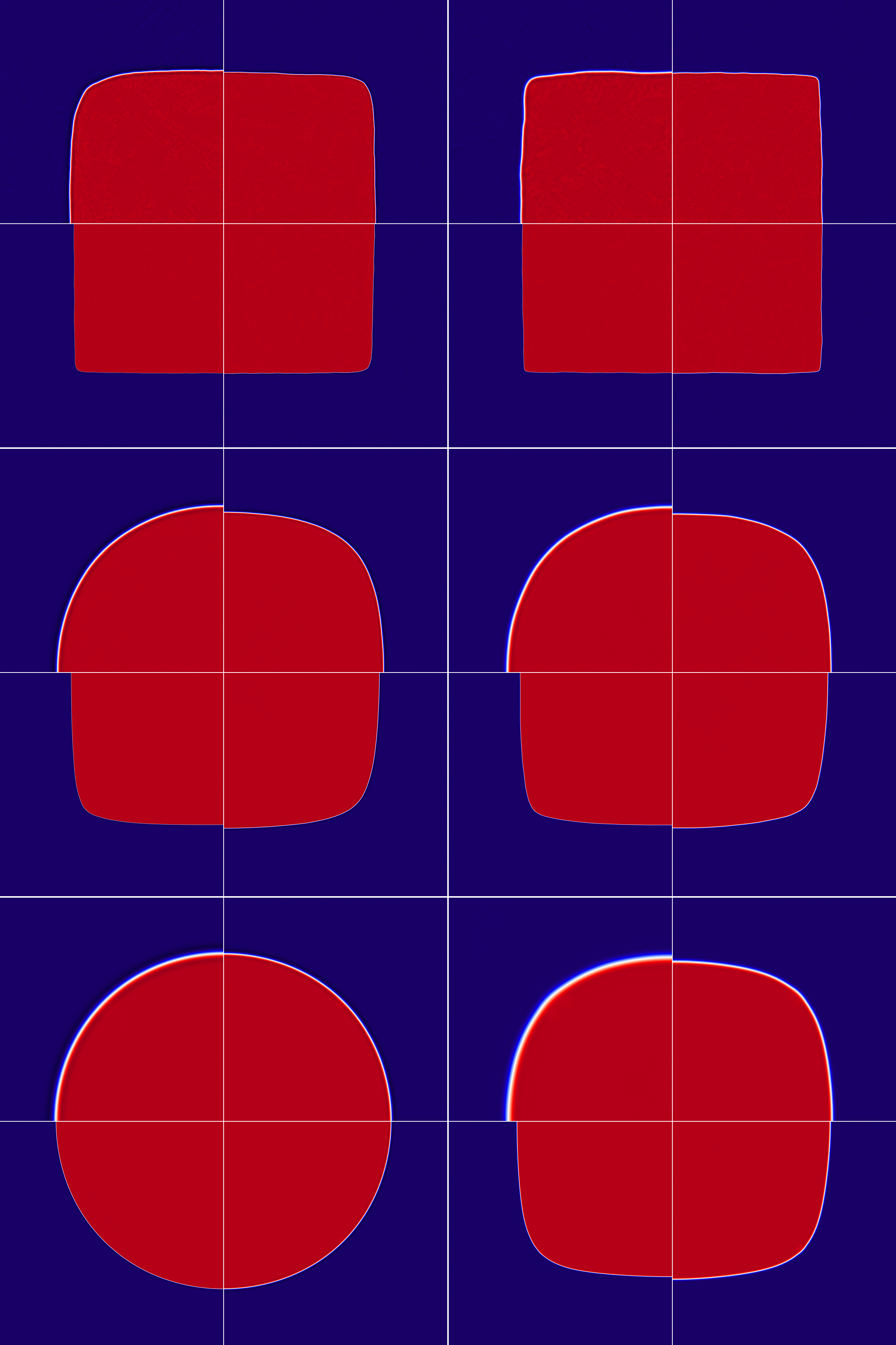}
\caption{Results of the hydrodynamic equilibrium test. Shown is the density, calculated by distributing the mass of each particle to each pixel according to the intersection between the pixel area and the particle's kernel using the SPH kernel function. The left column shows the results of the equal-mass test, while the right column shows the results of the equal-spacing test. In each block, the results are for $N=128$, $N=256$, $N=512$ and $N=1024$ ordered clockwise, starting from the top left panel. From top to bottom, the three boxes show the results with the cubic spline kernel with 32 neighbors, the Wendland C2 kernel with 100 neighbors and the Wendland C6 kernel with 400 neighbors.}
\label{fig:box_test_uncorrected_composite}
\end{figure}

This test is used to show the effect of the artificial surface tension introduced at material interfaces \citep{hopkinsGeneralClassLagrangian2013,saitohDENSITYINDEPENDENTFORMULATIONSMOOTHED2013,huSPHGalSmoothedParticle2014,hopkinsNewClassAccurate2015,wadsleyGasoline2ModernSmoothed2017,reinhardtBifurcationHistoryUranus2020}. In a periodic square domain ($-0.5\leq x,y,\leq 0.5$) of density $\rho_1$, a square region ($-0.25\leq x,y,\leq 0.25$) of high density $\rho_2$ is placed such that the two regions are in pressure equilibrium. For the low-density region, we use M-ANEOS forsterite \citep{stewartEquationStateModel2019} and for the high-density region, we use M-ANEOS iron \citep{stewartEquationStateModel2020a}. We perform this test in pseudo-2D with the low-density region sampled with $N^2\times16$ BCC cells, from which the central region is cut out and replaced with $(21N/16)^2\times21$ cells of equal-mass particles. This results in a density ratio between the two domains of $\rho_2/\rho_1=(21/16)^3 = \SI{2.261}{}$. We test the cubic spline kernel with 32 neighbors, the Wendland C2 kernel with 100 neighbors and the Wendland C6 kernel with 400 neighbors. We perform the same test also with an equal-spacing setup, where the masses of the particles in the two domains are changed such that the same density contrast is achieved. Figure~\ref{fig:box_test_uncorrected_composite} shows the results of both test series. In both the equal-mass and equal-spacing configuration, the test shows that with the cubic spline kernel, the square shape is very well conserved with only a small region around the corner being rounded. With the Wendland C2 kernel, the square shape gets more rounded, especially at low resolution, where the shape is more circular than square. With the Wendland C6 kernel with 400 neighbors, the square gets reshaped into a perfect circle in the equal-mass configuration. In the equal-spacing configuration, the result depends on the resolution. At low resolution we get a circle, while at higher resolution we get a square with strongly rounded corners. While nowhere near perfect, these results are much better than expected for a traditional SPH implementation, which we attribute mostly to the $\nabla h$ terms and the high resolution compared to prior work. Furthermore, the interface correction described in Section~\ref{sec:Interface_correction} does not alleviate this problem.

\subsection{Blob test}
\begin{figure*}[ht!]
\centering
\includegraphics[width=\linewidth]{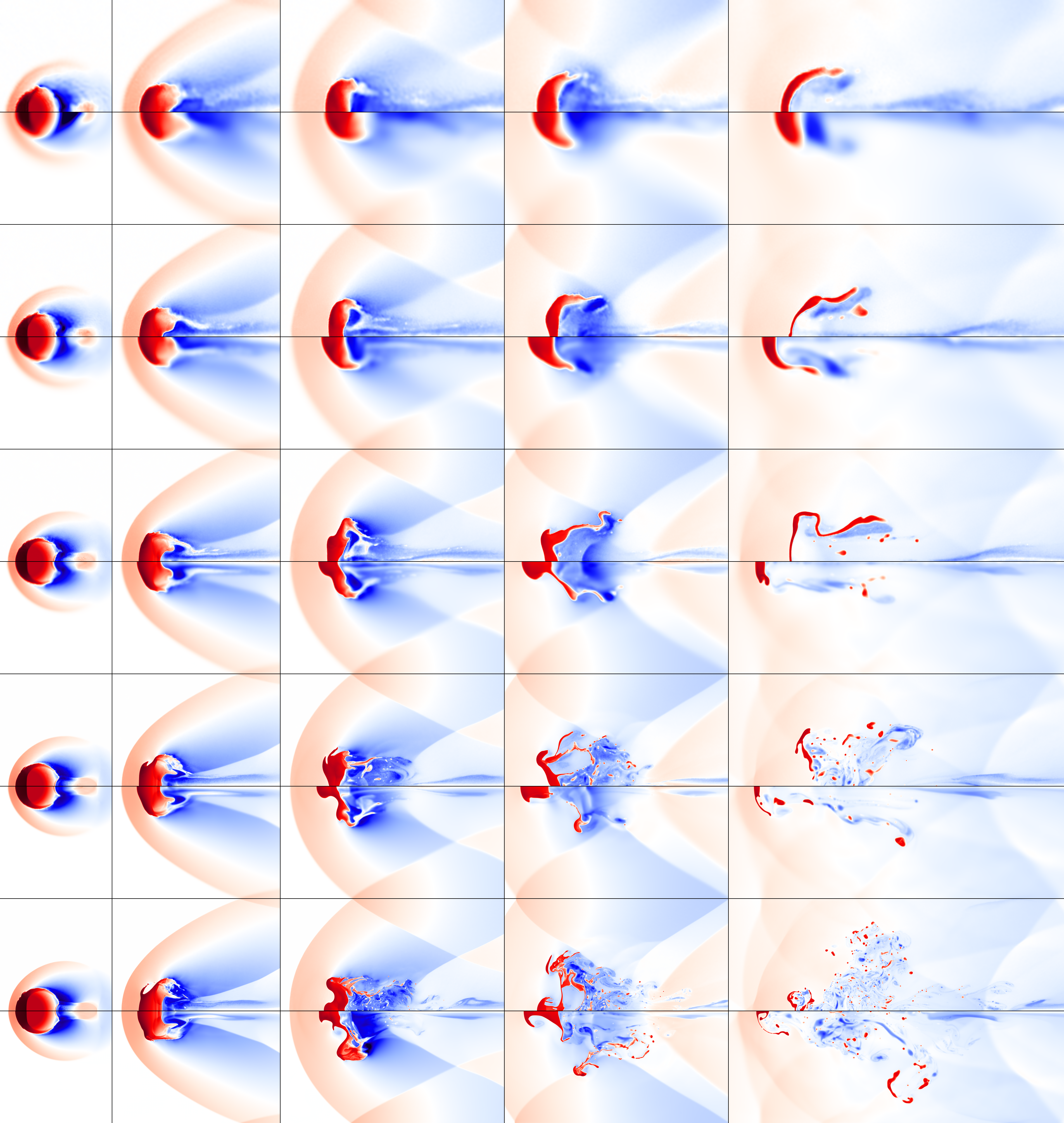}
\caption{Results from the Blob test. Each row shows a timeseries, from left to right: $0.25\,\tau_{KH}$, $1.0\,\tau_{KH}$, $1.75\,\tau_{KH}$, $2.5\,\tau_{KH}$ and $4.0\,\tau_{KH}$, where the top half of the panel is from the simulation using the cubic spline kernel with 32 neighbors and the bottom half is from the simulation using the Wendland C6 kernel with 400 neighbors. From top to bottom, the resolution increases, with the short sides of the domain being sampled with 64, 128, 256, 512 and 1024 BCC cells (see text for details).}
\label{fig:blob_test_composite}
\end{figure*}

\begin{figure}[ht!]
\centering
\includegraphics[width=\linewidth]{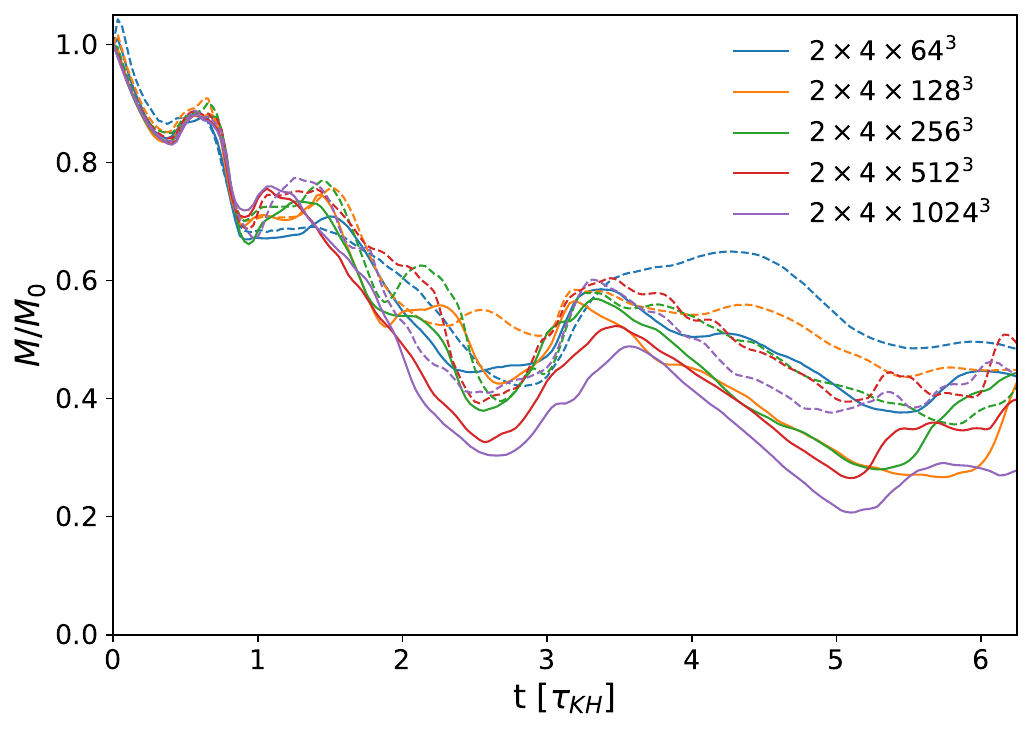}
\caption{The evolution of the cloud mass fraction in the blob test. The simulations with the cubic spline kernel and 32 neighbors are shown as solid lines, while the runs with the Wendland C6 kernel and 400 neighbors are shown as dashed lines.}
\label{fig:blobmass}
\end{figure}

One of the principal limitations of traditional SPH formulations lies in their treatment of multiphase fluid mixing. To examine this deficiency, \citet{agertzFundamentalDifferencesSPH2007} proposed the ``Blob'' test, which consists of a dense, initially stationary spherical cloud embedded within a uniform supersonic wind. The cloud is initialized in pressure equilibrium with the ambient medium. The expected behavior is that the blob undergoes progressive disruption as Kelvin-Helmholtz and Rayleigh-Taylor instabilities develop during its acceleration in the direction of the background flow. We create the initial conditions by sampling a $4\times1\times 1$ domain with $4N\times N^2$ BCC cells with a density of $\rho_w=1.0$ from which we cut out an $R=0.1$ sphere and fill the void with equal-mass particles in a BCC arrangement such that a density contrast of $\chi=10$ results. The particles in the cloud are assigned a temperature of $T_{cl} = \SI{10}{\kelvin}$, while those in the wind get a temperature of $T_{w}=\chi T_{cl} = \SI{100}{\kelvin}$. These initial conditions are then evolved for a while, applying a decreasing velocity damper and suppressing shock heating, to allow the particle positions, especially around the cloud, to relax. After that, the wind velocity of $\mathcal{M}=2.7$ is applied to the wind particles. The simulation is then run for $t=6.25\,\tau_{KH}$ where we use the definition $\tau_{KH}=3.2R_{cl}\sqrt{\chi}/v_w$ from \citet{agertzFundamentalDifferencesSPH2007}. Figure~\ref{fig:blob_test_composite} shows the resulting density distributions at different simulation times for different resolutions for both the cubic spline kernel with 32 neighbors and the Wendland C6 kernel with 400 neighbors. At low resolutions, we get similar results as \citet{agertzFundamentalDifferencesSPH2007}, the cloud does not break up, but gets stretched and deformed. With increasing resolution, we see more and more fragments getting ripped off the cloud, until at a resolution of $2\times4\times1024^3$, the cloud is shredded into many small pieces. But contrary to the results obtained with grid codes, the cloud material is never really ``mixed'' into the surrounding wind. This can also be seen in Figure~\ref{fig:blobmass}, which shows the evolution of the cloud mass (defined as all particles that satisfy $T<0.9\,T_w$ and $\rho>0.64\,\rho_{cl}$), as we never get below \SI{20}{\percent} of the initial cloud mass, even though increasing resolution reduces the cloud mass. This is partially due to the artificial surface tension (as discussed in Section~\ref{sec:Hydrostatic_equilibrium_test}), but also due to the conservation properties of SPH. As we ``pin'' entropy to each particle and only increase it in shocks, without introducing artificial entropy diffusion, the cloud will never dissolve into the wind in the same way as it does in grid codes \citep{wadsleyGasoline2ModernSmoothed2017}.

\subsection{Kelvin-Helmholtz instability}\label{sec:Kelvin-Helmholtz_instability}
\begin{figure}[ht!]
\centering
\includegraphics[width=\linewidth]{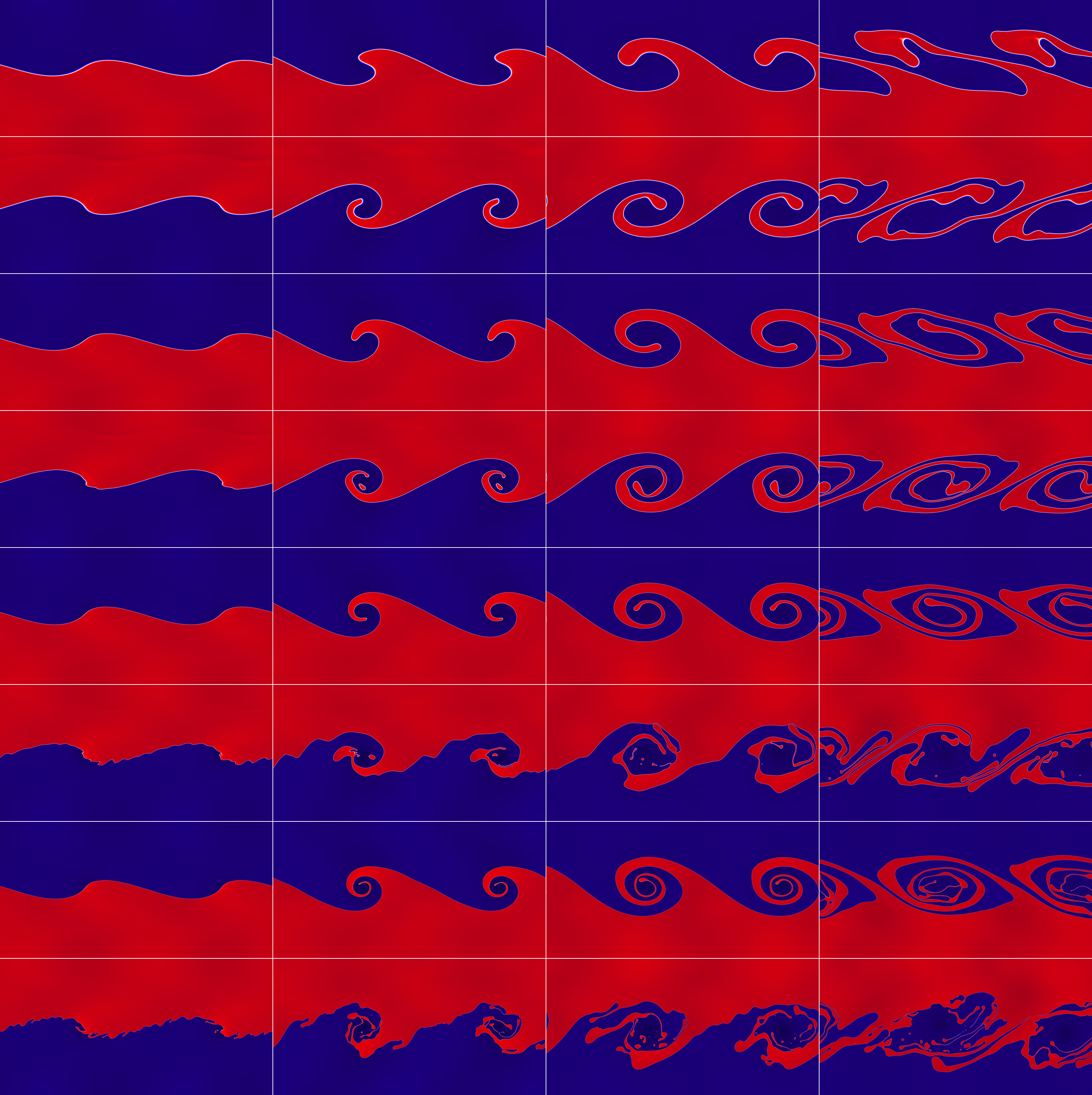}
\caption{Results of the Kelvin-Helmholtz instability test with a density ratio of $\chi=\SI{1.953}{}$ after (from left to right) $0.3\,\tau_{KH}$, $0.6\,\tau_{KH}$, $1.0\,\tau_{KH}$ and $2.0\,\tau_{KH}$. The top half of each row shows the results of the runs with $\alpha=1.0$ and $\beta=2.0$, while the lower half shows the runs with $\alpha=0.1$ and $\beta=0.0$. From top to bottom, the resolution increases, with the simulation $x$-axis in the low-density domain being sampled by 512, 1024, 2048, and 4096 BCC cells (see the text for details).}
\label{fig:kh_composite_density_ratio}
\end{figure}

\begin{figure}[ht!]
\centering
\includegraphics[width=\linewidth]{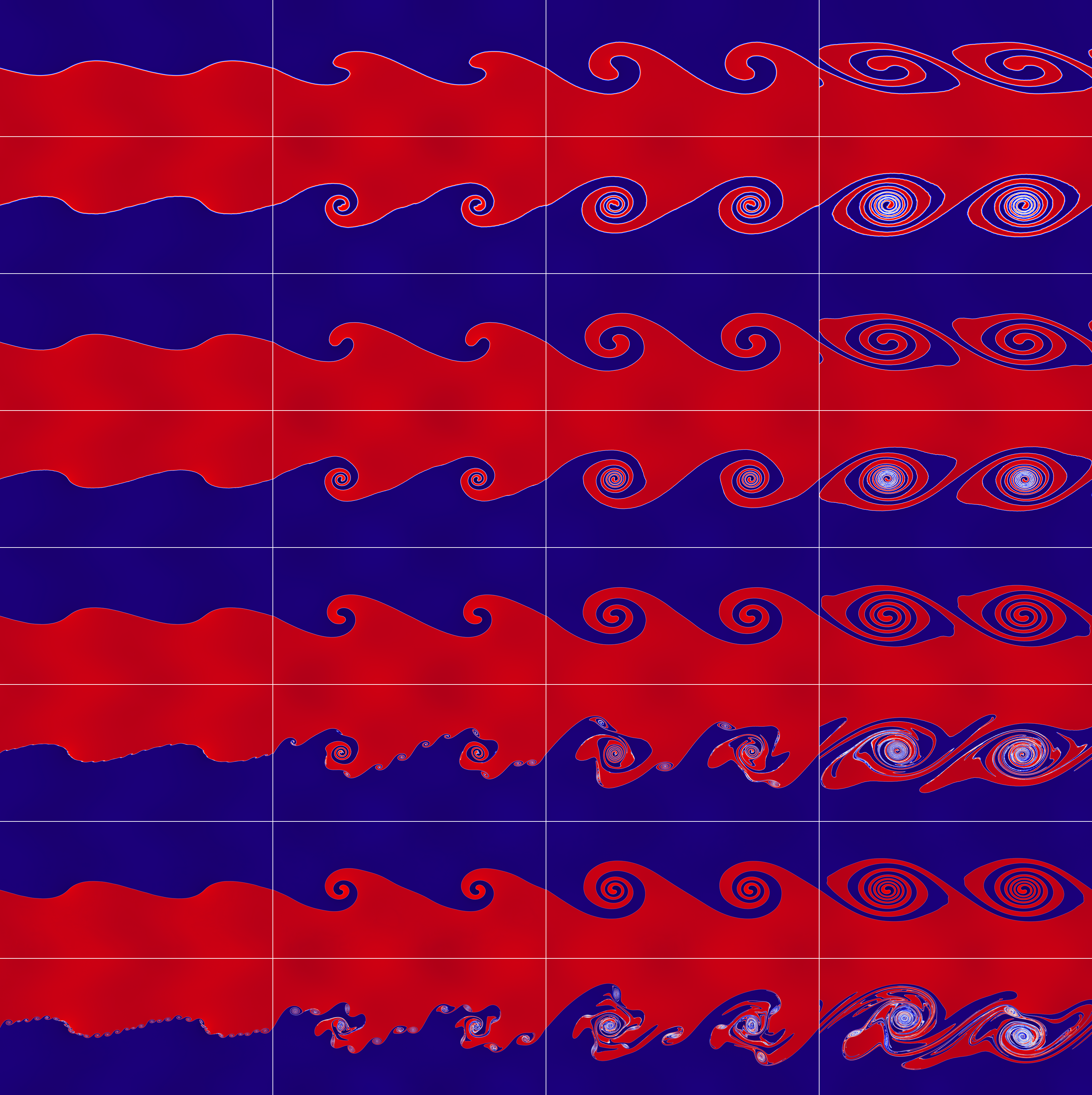}
\caption{Results of the Kelvin-Helmholtz instability test with a density ratio of $\chi=\SI{1}{}$ after (from left to right) $0.3\,\tau_{KH}$, $0.6\,\tau_{KH}$, $1.0\,\tau_{KH}$ and $2.0\,\tau_{KH}$. The top half of each row shows the results of the runs with $\alpha=1.0$ and $\beta=2.0$, while the lower half shows the runs with $\alpha=0.1$ and $\beta=0.0$. From top to bottom, the resolution increases, with the simulation $x$-axis being sampled by 512, 1024, 2048, and 4096 BCC cells (see the text for details).}
\label{fig:kh_composite_equal_density}
\end{figure}

The Kelvin-Helmholtz instability provides a way to test how the artificial surface tension and the artificial viscosity suppress the mixing by fluid instabilities. We follow \citet{readResolvingMixingSmoothed2010} to create the initial conditions, but as we use a quasi-2D setup, we are limited in the choice of the density contrast. We choose 20 BCC cells for the high-density domain and 16 cells for the low-density domain, resulting in a density contrast of $\chi=\rho_{1}/\rho_{2}=(20/16)^3 = \SI{1.953}{}$, but we also run the test without a density contrast, i.e. 16 cells in both domains. In a periodic square domain ($-0.5\leq x,y,\leq 0.5$), we create two grids with equal-mass particles, one with $\rho_{1}$, the other with $\rho_{2}$. We then remove particles such that the central slab with $\left\vert y\right\vert < 0.25$ consists of the material with $\rho_{1}$, and the remaining domain of the material with $\rho_{2}$. The whole volume is initialized with uniform pressure such that $\rho_{1}/\rho_{2} = T_{2}/T_{1}=c_{2}^2/c_{1}^2$. The two layers are given constant and opposing shearing velocities with the outer layer having $\mathcal{M}_{2} = -v_{2}/c_{2}=0.11$ and the central layer moving at $\mathcal{M}_{1}=\mathcal{M}_{2}\sqrt{\chi}$. With the relative velocity $v = v_1-v_2$ we get the Kelvin-Helmholtz growth time as $\tau_{KH}=\frac{(\rho_1+\rho_2)\lambda}{\sqrt{\rho_1\rho_2}v}$. We seed the instability by applying a velocity perturbation at the two boundaries of the form \citep{readResolvingMixingSmoothed2010}

\begin{align}
v_y&=\delta v_y \left[\sin(2\pi(x+\lambda/2)/\lambda)\exp(-(-10(y-0.25))^2)\right.\nonumber\\
&\left.+\sin(2\pi x/\lambda)\exp(-(10(y+0.25))^2)\right]
\end{align}

\noindent where the perturbation amplitude is $\delta v_y=v/8$ and the wavelength is $\lambda=0.5$. We run the tests for both density ratios with the same two choices for the artificial viscosity parameters we used in the Gresho-Chan vortex test (see Section~\ref{sec:Gresho-Chan_vortex}), the combination capable of capturing shocks ($\alpha=1.0$ and $\beta=2.0$) and a low-viscosity combination ($\alpha=0.1$ and $\beta=0.0$) that is not able to capture shocks. Figure~\ref{fig:kh_composite_density_ratio} shows the results for the test with a density ratio of $\chi=\SI{1.953}{}$ between the two layers, while Figure~\ref{fig:kh_composite_equal_density} shows the results for the test with a density ratio of $\chi=\SI{1}{}$. With the high viscosity, the development of the instability is strongly suppressed, but at higher resolutions, the characteristic rolls develop for both values of $\chi$. With low viscosity, the rolls develop faster, especially in the $\chi=1$ case. Above a certain resolution, the flow starts to develop instabilities at smaller wavelengths than the one that is seeded. In the simulations with $\chi=\SI{1.953}{}$, this leads to the destruction of the rolls as the secondary rolls that should develop are suppressed by the surface tension. In the simulations with $\chi=1$, the rolls of the secondary instabilities are allowed to develop and the primary instability rolls up the smaller ones into its structure. The two fluids are nevertheless kept separate and no mixing occurs. It becomes thus clear that the artificial viscosity and the artificial surface tension at the interface suppress the development of the instability in unique ways, but increasing the resolution improves the results significantly.

\section{Performance and scaling}\label{sec:Scaling_tests}
In this section, we want to demonstrate the excellent performance and scaling characteristics of the SPH implementation in \texttt{pkdgrav3}. All of the tests in this section were performed on the Piz Daint hybrid CPU-GPU supercomputer at CSCS in Lugano, Switzerland. We used up to 1024 nodes of the more than 5700 nodes of the machine. Each Cray XC50 node consists of one 12 core Intel Xeon E5-2690 v3 CPU with 64 GB RAM and one Nvidia Tesla P100 GPU with 16 GB vRAM.

\subsection{Method scaling}\label{sec:Method_scaling}
\begin{figure}[ht!]
\centering
\includegraphics[width=\linewidth]{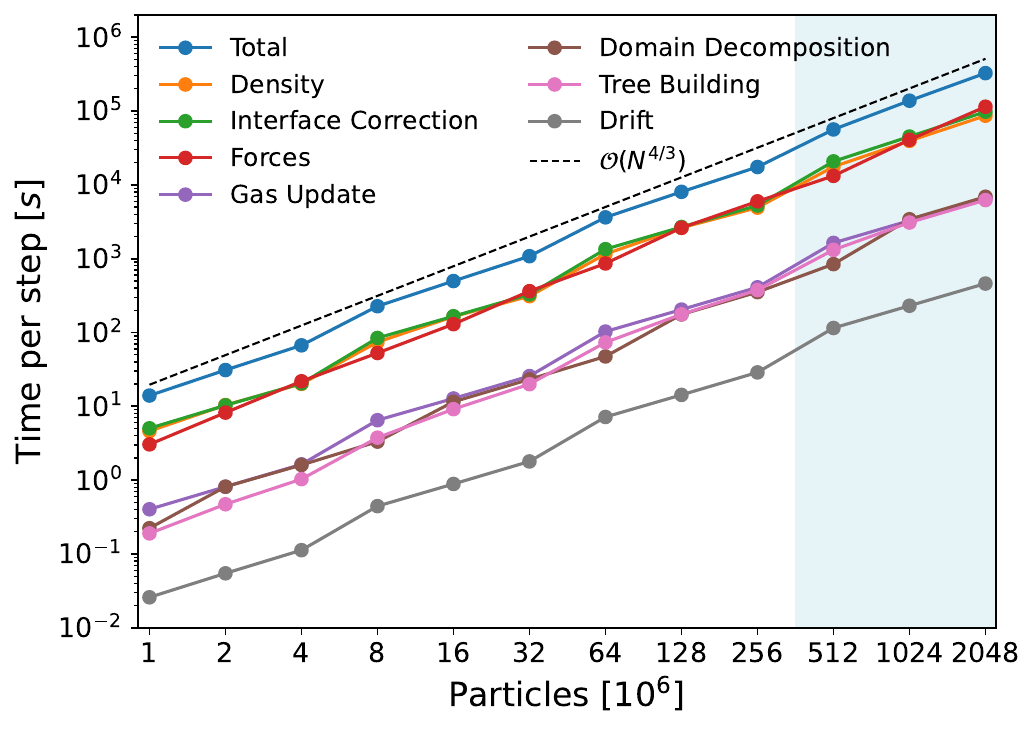}
\caption{Scaling of time per step on a single node. The total time per step closely follows the expected $\mathcal{O}\bigl(N^{4/3}\bigr)$ scaling. The values in the blue shaded region are extrapolated from runs with more than one node, because they did not fit into memory on one node.}
\label{fig:method_scaling}
\end{figure}

First, we want to measure how the code scales on a single node, with increasing number of particles. Thanks to the fact that both the gravity as well as the neighbor finding are $\mathcal{O}(N)$ (see Sections~\ref{sec:Gravity} and~\ref{sec:Neighbor_search}), we expect linear scaling of the runtime for a single substep. The timestep size needed to satisfy the CFL condition is proportional to the kernel size which scales with $N^{-\frac{1}{3}}$. This means that the number of substeps in a global step done by the code scales with $N^\frac{1}{3}$. Thus, the expected scaling of a global step with as many substeps as necessary to satisfy the CFL condition is $\mathcal{O}\bigl(N^{4/3}\bigr)$. The benchmark case we use is a Mars-sized body consisting of an M-ANEOS iron \citep{stewartEquationStateModel2020a} core and a M-ANEOS forsterite \citep{stewartEquationStateModel2019} mantle. It is sampled with various numbers of particles between 1 million and 2 billion. These models are then evolved for one global step, the size of which is chosen such that the test with 1 million particles needs 8 substeps. Figure~\ref{fig:method_scaling} shows the timing measurements for all tested particle counts. The last three data points are extrapolated from runs with more nodes as the simulation would not fit into memory on one node (which is limited to 64 GB). The total time per step closely follows the expected scaling over the full range of tested particle counts. We also record the time taken by the different operations. The total step time is dominated by the three operations that do tree walks: the density calculation, the interface correction and the force calculation (which includes both the gravity and the SPH forces), each contributing roughly one third of the run time. These also contain most of the calls to the EOS, which can not be timed separately, as they happen interleaved with the tree walk. After the interface correction, a separate fluid value update with calls to the EOS is performed, which is timed separately. This creates only a subdominant contribution to the step time, likewise for the domain decomposition and the tree build. The smallest contribution comes from the drift.

\subsection{Strong scaling}
\begin{figure}[ht!]
\centering
\includegraphics[width=\linewidth]{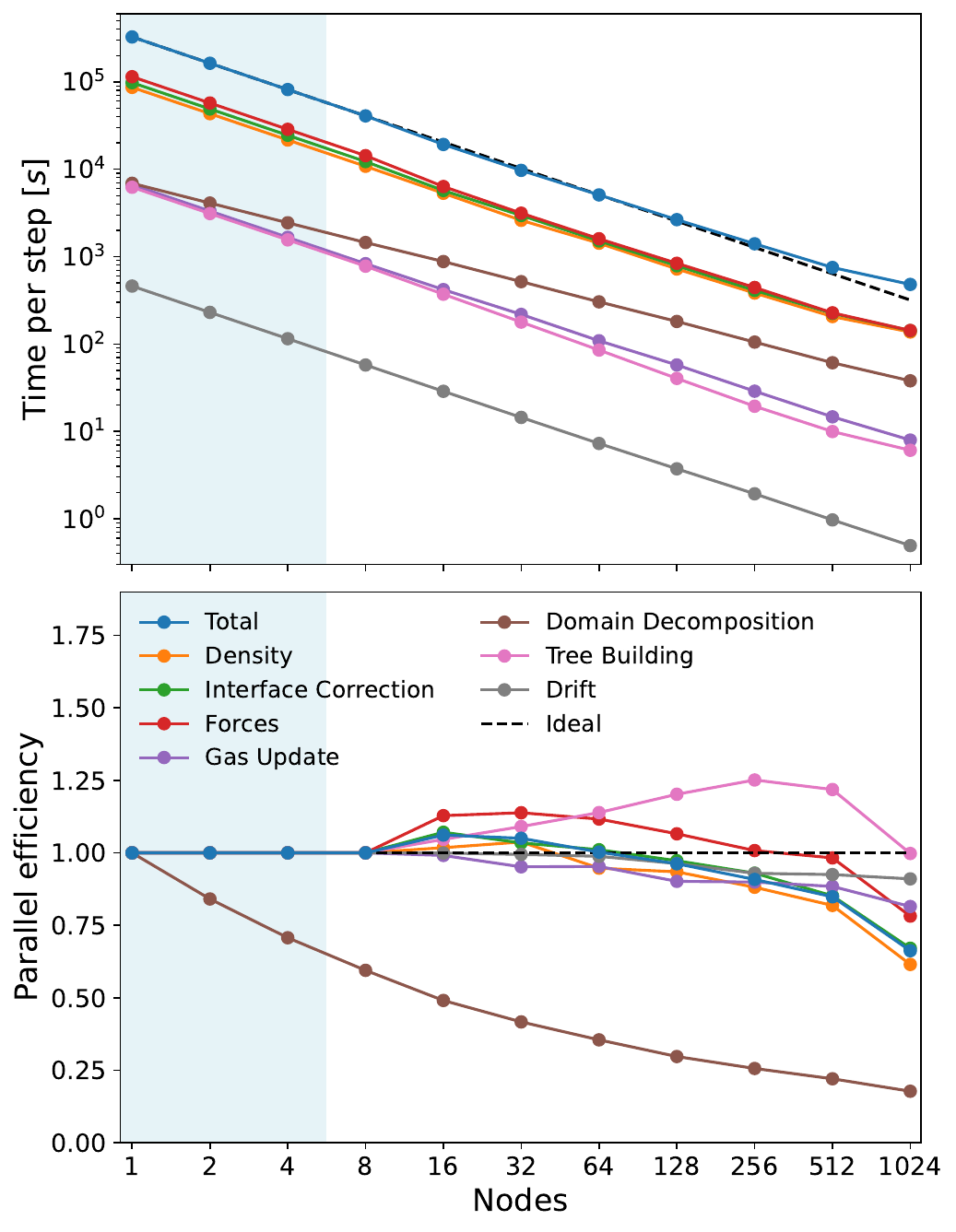}
\caption{Strong scaling results with 2048 million particles. The top panel shows the time per step and the bottom panel shows the parallel efficiency. The values for 1, 2 and 4 nodes in the blue shaded region are extrapolated from the result for 8 nodes, as the simulation would not fit into memory at those node counts.}
\label{fig:strong_scaling_2048}
\end{figure}

\begin{figure}[ht!]
\centering
\includegraphics[width=\linewidth]{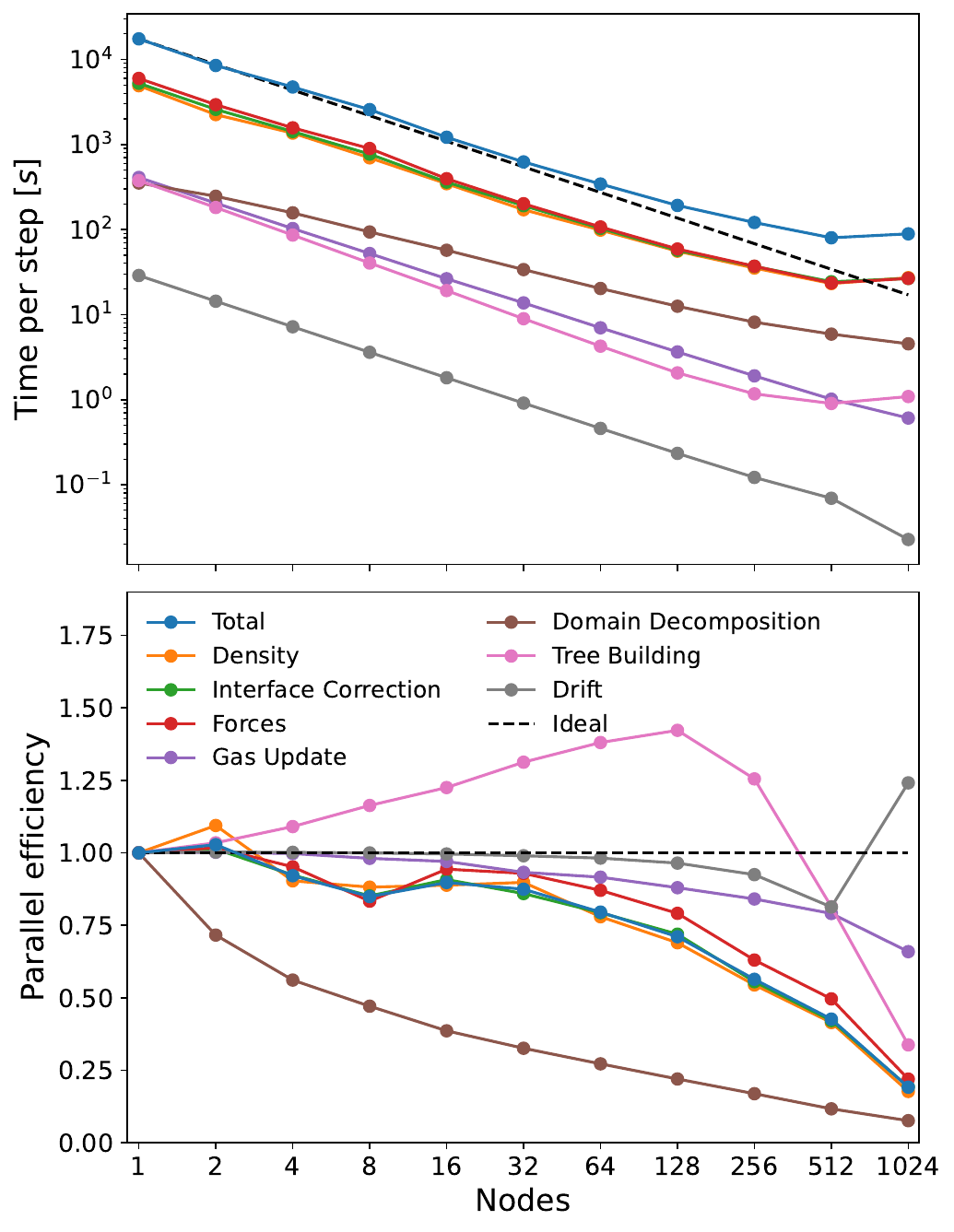}
\caption{Strong scaling results with 256 million particles. The top panel shows the time per step and the bottom panel shows the parallel efficiency.}
\label{fig:strong_scaling}
\end{figure}

For the strong scaling tests, we use the 2048 million and 256 million particle models of the same Mars-sized body as in Section~\ref{sec:Method_scaling} and run them for one global step with several different node counts between 1 and 1024. Figure~\ref{fig:strong_scaling_2048} shows the results for the model with 2048 million particles. The values for 1, 2 and 4 nodes are extrapolated from the results with 8 nodes, assuming perfect scaling, as it does not fit into memory at those node counts. The results show very good strong scaling with parallel efficiency above \SI{80}{\percent} until 512 nodes, where it then decreases to a still respectable \SI{66}{\percent} at 1024 nodes with a speedup of \SI{678}{}, considering that there are only \SI{180000}{} particles per thread left at that point. We can also see that all operations follow a linear scaling with $N_{nodes}^{-1}$, with the exception of the domain decomposition which scales slightly worse with $N^{-\frac{3}{4}}$, but it stays subdominant over the whole range. Figure~\ref{fig:strong_scaling} shows the results for the model with 256 million particles. We see the same trends as for the higher resolution test, but we get parallel efficiency above \SI{80}{\percent} only for up to 64 nodes, above which efficiency steadily declines to \SI{56}{\percent} at 256 nodes and only \SI{20}{\percent} at 1024 nodes. The maximum speedup of \SI{218}{} is achieved at 512 nodes where we have \SI{42}{\percent} efficiency. We performed these tests also for all other models used in Section~\ref{sec:Method_scaling} and we find that in no case do we get speedups below \SI{1.0}{}. As expected, the lowest value of parallel efficiency (\SI{0.68}{\percent}) is achieved when distributing 1 million particles onto 1024 nodes. In this case, we still get a speedup of 7, even though there are only roughly 90 particles per processor, less than a group size worth.

\subsection{Weak scaling}
\begin{figure}[ht!]
\centering
\includegraphics[width=\linewidth]{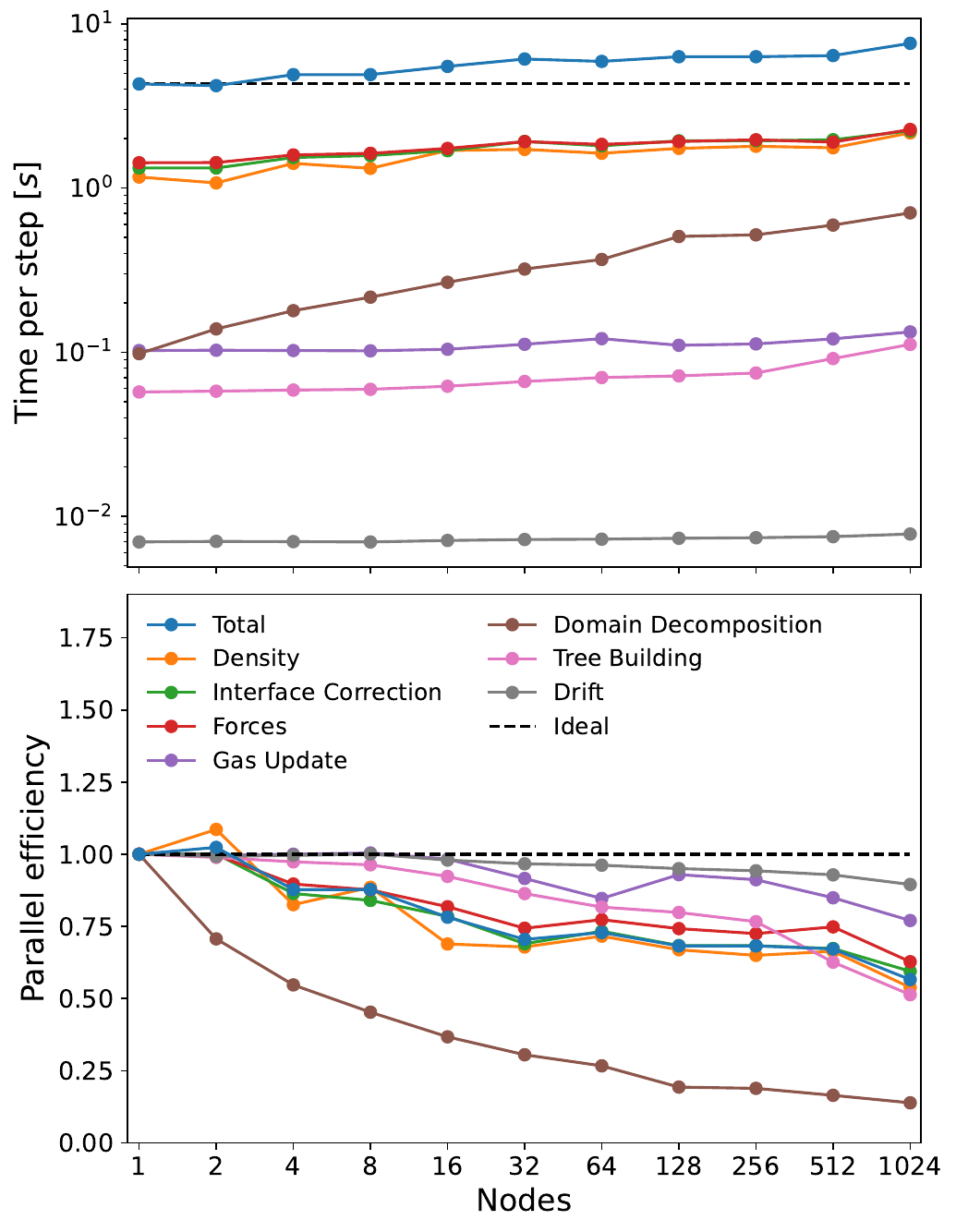}
\caption{Weak scaling results with 2 million particles per node. The top panel shows the time per step and the bottom panel shows the parallel efficiency.}
\label{fig:weak_scaling}
\end{figure}

\begin{figure}[ht!]
\centering
\includegraphics[width=\linewidth]{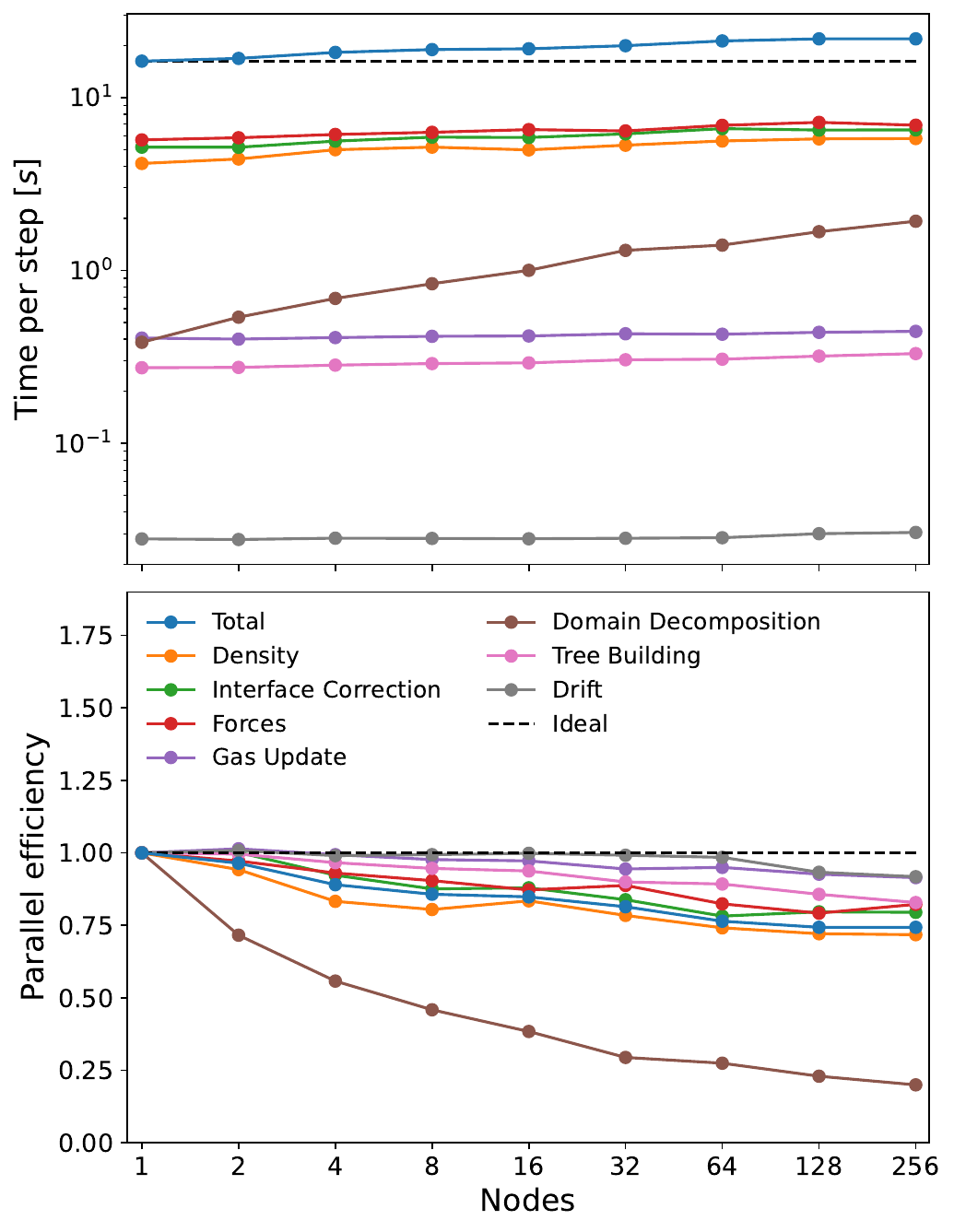}
\caption{Weak scaling results with 8 million particles per node. The top panel shows the time per step and the bottom panel shows the parallel efficiency.}
\label{fig:weak_scaling_256}
\end{figure}

We also present weak scaling results. For this, we use the same models from Section~\ref{sec:Method_scaling} again, but we choose the global step size such that it satisfies the CFL condition for all resolutions directly, removing the need for the code to do substeps and thus also the $\mathcal{O}\bigl(N^{1/3}\bigr)$ scaling introduced by the substepping. To reduce the noise in the results, we run the test for 10 steps and average the resulting times. We show weak scaling results with 2 million particles per node in Figure~\ref{fig:weak_scaling}. This corresponds to the highest resolution model with 2048 million particles being run on 1024 nodes. But as shown in Figure~\ref{fig:strong_scaling_2048}, at this node count, the result has already visibly departed from perfect scaling. A production simulation would thus never be run in this configuration, but rather with a more realistic 256 nodes. Thus, we also show the corresponding weak scaling results with 8 million particles per node in Figure~\ref{fig:weak_scaling_256}. Both figures show that all operations scale reasonably well, with the results with 2 million particles per node staying above \SI{50}{\percent} parallel efficiency and the results with 8 million particles per node above \SI{75}{\percent}. The exception is in both cases the domain decomposition, which scales significantly worse, but because it stays very subdominant over the whole range, it is acceptable.

\section{Conclusions}\label{sec:Conclusions}
We present a high-performance implementation of Smoothed Particle Hydrodynamics (SPH) with a novel approach to neighbor finding that couples optimally with the Fast Multipole Method (FMM) gravity solver in \texttt{pkdgrav3}. The vectorization of particle-particle interactions allows efficient execution on both CPU-only and hybrid CPU/GPU architectures. This design gives \texttt{pkdgrav3} excellent performance and scaling characteristics on systems ranging from laptops to modern high-performance computing (HPC) systems.

The resulting performance enables new types of studies in two complementary ways. First, it allows simulations with unprecedented particle counts, making it possible to resolve low-mass planetary features such as crusts, oceans, and tenuous atmospheres. Second, due to the code being highly efficient even at modest resolutions, large parameter spaces can be explored with several million particles which was previously limited to low resolution models \citep[e.g.,][]{timpeSystematicSurveyMoonforming2023,meierSystematicSurveyMoonforming2024,emsenhuberNewDatabaseGiant2024}. \texttt{pkdgrav3} has already been used in both capacities in recent peer-reviewed studies:

\begin{itemize}
    \item In \citet{matzkevichOutcomeCollisionsGaseous2024}, \texttt{pkdgrav3} was used to explore the parameter space of collisions between clumps formed by disk instability using simulations with up to 2 million particles, each run on either one Eiger.Alps node with 128 CPU cores or 4 Piz Daint CPU nodes with 36 CPU cores each. The simulation with 1.5 million particles shown in their Figure~6, which results in dynamic collapse where the density suddenly increases by $\sim 6$ orders of magnitude, was even run on a workstation with only 16 CPU cores and no GPU acceleration.
    \item In \citet{bussmannPossibilityGiantImpact2025} \texttt{pkdgrav3} was again used to perform a parameter space exploration, here with simulations with $\sim$ 800k particles of giant impacts on Venus, most of them run on a single Piz Daint GPU node.
    \item In the appendix of \citet{meierSystematicSurveyMoonforming2024} we presented simulations with 10 million particles investigating the direct formation of the Moon proposed in \citet{kegerreisImmediateOriginMoon2022}. For the proposed scenario, at least 10 million particles are necessary to resolve the angular momentum transport in the orbiting material.
    \item Finally, in \citet{meierOriginJupitersFuzzy2025}, \texttt{pkdgrav3} was used to study whether Jupiter's dilute core could be the result of a giant impact. Since core disruption and possible mixing require extremely high resolution, the flagship simulation was performed with 2.1 billion particles, which marks, to our knowledge, the highest resolution giant impact simulation to date. This simulation was run on 256 Piz Daint GPU nodes.
\end{itemize}

These examples illustrate the flexibility and scalability of \texttt{pkdgrav3}, demonstrating that it can efficiently address problems across an exceptionally wide range of physical and computational scales.

In future work, we plan several extensions and improvements to further enhance both the physical fidelity and numerical performance of the code. We will investigate which artificial viscosity limiting scheme to incorporate to reduce the effect on angular momentum transport without compromising shock capturing. Additional physics modules, such as strength and failure models (to be presented in Paper~II), subgrid treatments for stellar and feedback processes, and radiative cooling, will broaden the range of astrophysical applications. Finally, we aim to enable double-precision computations and I/O on demand. While this will increase data volume and may reduce raw performance by up to a factor of two, preliminary tests suggest that communication overhead can partially mitigate this effect. As a result of current and ongoing development \texttt{pkdgrav3} can provide both unprecedented resolution and physical realism bringing shock physics modeling to the next level.

\begin{acknowledgments}
We thank the anonymous reviewer for valuable suggestions and comments that substantially helped to improve the paper. This work has been carried out within the framework of the National Centre of Competence in Research PlanetS supported by the Swiss National Science Foundation under grants 51NF40\_182901 and 51NF40\_205606. The authors acknowledge the financial support of the SNSF. TM acknowledges support from the University of Zurich through a Candoc grant. We acknowledge access to Eiger.Alps at the Swiss National Supercomputing Centre, Switzerland under the University of Zurich's share with the project ID UZH4. This work was supported by a grant from the Swiss National Supercomputing Centre (CSCS) under project IDs S1285 and LP87 on Piz Daint and Daint.Alps.
\end{acknowledgments}

\begin{contribution}

T.M. implemented the SPH module in \texttt{pkdgrav3}, performed the hydrodynamics and scaling tests, and was responsible for writing and submitting the manuscript. D.P. is the primary maintainer of the \texttt{pkdgrav3} repository, assisted T.M. with the implementation, and contributed to manuscript editing. C.R. provided expertise on details of the SPH implementation and planetary scale impacts, secured the computational resources for the validation runs, and edited the manuscript. J.S. conceived the initial neighbor-finding concept, obtained the funding that supported T.M., and edited the manuscript.


\end{contribution}

%
\facilities{Swiss National Supercomputing Centre (Piz Daint, Eiger.Alps, Daint.Alps)}

\software{pkdgrav3 \citep{potterPKDGRAV3TrillionParticle2017},
          ballic \citep{reinhardtNumericalAspectsGiant2017},
          eoslib \citep{meierEOSlib2021,meierANEOSmaterial2021},
          tipsy \citep{n-bodyshopTIPSYCodeDisplay2011},
          numpy \citep{harrisArrayProgrammingNumPy2020},
          scipy \citep{virtanenSciPy10Fundamental2020},
          matplotlib \citep{hunterMatplotlib2DGraphics2007},
          GNU parallel \citep{tangeGNUParallelCommandline2011}
          }




\bibliography{main}{}
\bibliographystyle{aasjournalv7}



\end{document}